\documentclass[twocolumn,apj,numberedappendix,twocolappendix,iop]{openjournal}

\usepackage[T1]{fontenc}

\usepackage{graphicx}	
\usepackage{amsmath}	
\usepackage{newtxtext,newtxmath}    
\usepackage[dvipsnames]{xcolor}  
\usepackage[colorlinks,linkcolor=blue,citecolor=blue,urlcolor=blue ]{hyperref}
\usepackage{amssymb}	
\usepackage{dsfont}
\usepackage{ulem}
\usepackage{soul}
\usepackage{enumitem}
\usepackage{orcidlink}
\usepackage{hyperref}
\usepackage{multirow}

\usepackage{natbib}

\usepackage{fontawesome}
\usepackage{lineno}

\usepackage{macros_vdbosch} 

\defcitealias{C43}{C43}

\DeclareRobustCommand{\VAN}[3]{#2}
\let\VANthebibliography\thebibliography
\def\thebibliography{\DeclareRobustCommand{\VAN}[3]{##3}\VANthebibliography}

\graphicspath{{./}{figures/}}

\begin{document}


\title{Not all cores are equal:\\ 
Phase-space origins of dynamical friction, stalling and buoyancy}

\shorttitle{Not all cores are equal}
\shortauthors{Dattathri et al.}

\author{Shashank Dattathri$^1$\orcidlink{0000-0002-7941-1149}}
\author{Frank C. van den Bosch$^1$\orcidlink{0000-0003-3236-2068}}   
\author{Uddipan Banik$^{2,3,4}$\orcidlink{0000-0002-9059-381X}}
\author{Martin Weinberg$^{5}$\orcidlink{0000-0003-2660-2889}}
\author{Priyamvada Natarajan$^{1,6}$\orcidlink{0000-0002-5554-8896}}
\author{Zhaozhou Li$^{7,8}$\orcidlink{0000-0001-7890-4964}}
\author{Avishai Dekel$^{8,9}$\orcidlink{0000-0003-4174-0374}}

\affiliation{$^1$Department of Astronomy, Yale University, PO. Box 208101, New Haven, CT 06520-8101}
\affiliation{$^2$Department of Astrophysical Sciences, Princeton University, 112 Nassau Street, Princeton, NJ 08540, USA}
\affiliation{$^3$Institute for Advanced Study, Einstein Drive, Princeton, NJ 08540, USA}
\affiliation{$^4$Perimeter Institute for Theoretical Physics, 31 Caroline Street N., Waterloo, Ontario, N2L 2Y5, Canada}
\affiliation{$^5$Department of Astronomy, University of Massachusetts, Amherst, MA 01003-9305, USA}
\affiliation{$^6$ Department of Physiscs, Yale University, P.O. Box 208102, New Haven, CT 06520-8102}
\affiliation{$^7$School of Astronomy and Space Science, Nanjing University, Nanjing, Jiangsu 210093, China}
\affiliation{$^8$Racah Institute of Physics, The Hebrew University, Jerusalem 91904, Israel}
\affiliation{$^9$SCIPP, University of California, Santa Cruz, CA 95064, USA}

\email{shashank.dattathri@yale.edu}  

\label{firstpage}


\begin{abstract}
  Dynamical friction governs the orbital decay of massive perturbers within galaxies and dark matter halos, yet its standard Chandrasekhar formulation fails in systems with cores of (roughly) constant density, where inspiral can halt or even reverse, phenomena known respectively as core stalling and dynamical buoyancy. Although these effects have been observed in simulations, the conditions under which they arise remain unclear. Using high-resolution N-body simulations and analytic insights from kinetic theory, we systematically explore the physical origin of these effects. We demonstrate that the overall distribution function (DF) of the host, not just its central density gradient, determines the efficiency and direction of dynamical friction. Core stalling arises when the perturber encounters a plateau in the DF, either pre-existing or dynamically created through its own inspiral, while buoyancy emerges in systems whose DFs possess an inflection that drives an unstable dipole mode. We show that double power-law density profiles with rapid outer-to-inner slope transitions naturally produce such DF features, which is why structurally similar cores can yield radically different dynamical outcomes. Our results provide a unified framework linking the phase-space structure of galaxies to the fate of embedded massive objects, with direct implications for off-center AGN, the dynamics of nuclear star clusters, and the stalled coalescence of black holes in dwarf galaxies and massive ellipticals. 
\end{abstract}



\section{Introduction}
\label{sec:intro}

Dynamical friction is an essential ingredient of hierarchical structure and galaxy formation. On large scales, it is the dominant process that enables galaxies inside dark matter halos to merge \citep{White1978, MoVdBWhite}. Inside galaxies, it is assumed to be responsible for the formation of nuclear star clusters from the merging of globular clusters that sink to the galactic center \citep{Tremaine1975, CapuzzoDolcetta2008}, and for supermassive black holes (SMBHs) to inspiral towards each other until gravitational wave radiation enables their merger \citep{Milosavljevic2001}. The study of dynamical friction originates from the seminal paper by \citet[][hereafter \citetalias{C43}]{C43}, which showed that a massive perturber moving through an infinite, homogeneous, and isotropic medium experiences a drag force, given by 
\begin{equation}\label{eq:c43}
\mathbf{F}_{\rm DF}=-\frac{4\pi G^2 M_\rmP^2 }{v^2}\ln{\Lambda} \rho(<v) \frac{\mathbf{v}}{v} \,.
\end{equation}
Here, $M_\rmP$ is the mass of the perturber and $\mathbf{v}$ is its velocity, $\rho(<v)$ is the local density of stars with a speed less than $v=\left| \mathbf{v} \right|$, and $\ln{\Lambda}=\ln{b_{\rm max}/b_{\rm min}}$ is the Coulomb logarithm, with $b_{\rm max}$ and $b_{\rm min}$ the maximum and minimum impact parameters, respectively. 

Despite the oversimplified assumptions made in the derivation of equation~(\ref{eq:c43}), in many cases it has been shown to be fairly accurate in predicting the inspiral of massive perturbers \citep[e.g.][]{Lin1983,vdb1999}. However, there are also cases where it clearly fails. A prominent example is the orbital decay of perturbers in systems with a constant-density core. Equation~(\ref{eq:c43}) predicts that the perturber should continue to sink toward the center. However, \citet{Weinberg1986} predicted analytically that the perturber would stall when it comes close to the core radius (see also \citealt{Kalnajs1972}), which was subsequently verified with numerical simulations \citep{Read2006, Inoue2011, Cole2012, Chowdhury2019}. This phenomenon is known as ``core stalling''. Even more intriguing is the fact that, as first demonstrated by \citet{Cole2012}, perturbers that are initially placed within a core start to move outward, a phenomenon that has become known as ``dynamical buoyancy''.

Cores seem to be fairly ubiquitous among galaxies. In particular, rotation curves of dwarf and low-surface-brightness galaxies often reveal that their dark matter halos are cored \citep[see] [for a historical overview]{deBlok2010}. Dark matter cores have also been kinematically inferred in the halos of massive star-forming disks at redshifts $z \sim 1-3$ \citep[][]{Genzel2020, Price2021}. Although dark matter halos in the $\Lambda$CDM paradigm are predicted to form with central cusps, there are various processes related to the formation and evolution of galaxies that can transform such a dark matter cusp into a core. These include explosive gas outflows driven by supernova or AGN feedback \citep[e.g.,][]{Ogiya2011,Pontzen2012, Teyssier2013, Dekel2021, Li2023, Koudmani2024}, the tidal disruption of halos with radial anisotropy \citep[][]{Chiang.etal.24}, the dynamical heating of central cusps due to the inspiral of a massive object \citep[][]{ElZant2001, Goerdt2010, Cole2011, Ogiya2022}, and bar-driven halo evolution \citep{Weinberg2002}. Dark matter cores are also a natural prediction of several alternative dark matter models, such as Self-Interacting Dark Matter \citep[SIDM;][]{Spergel2000, Kochanek2000} or Fuzzy Dark Matter \citep[FDM;][]{Schive2014}. Cores are also common in the {\it stellar} density profiles of dwarf spheroidals \citep[][]{Moskowitz2020} and of the most massive ellipticals \citep{Ferrarese.etal.94, Gebhardt.etal.96}. For the latter, these cores are believed to be the outcome of SMBH binary scouring \citep{Begelman.etal.80, Quinlan1997, Partmann2024} and/or the back-and-forth sloshing of a recoiling SMBH following a merger \citep{Nasim2021, Rawlings2025}. 

Hence, cores are not observational exceptions, but rather a fairly natural product of structure formation. This implies that core stalling and dynamical buoyancy might have profound implications for galaxy evolution. In fact, core stalling has already been invoked to explain the presence of multiple nuclei or globular clusters that, in the absence of stalling, would have coalesced in a short time \citep{Goerdt.etal.06, Cole2012, Chowdhury2019, Boldrini2019, Meadows2020, SanchezSalcedo2022}. Core stalling and buoyancy may also explain the relatively large fractions of dwarf galaxies with AGN and/or nuclear star clusters that are offset from the photometric center \citep{Bingelli.etal.00, Reines2020, Mezcua2024}\footnote{Our analysis focus on single perturbations, and therefore does not take into account perturber-perturber interactions such as globular clusters scattering off of each other \citep{Chowdhury2019,Poulain2025}, black hole-black hole interactions \citep{Partmann2024}, and black hole-stellar clump scattering \citep{Ma2021}. The purpose of this paper is to show that inspiral can be complicated even under the ``best case" scenario where a perturber is sinking in a smooth, symmetric host system.}. Arguably, the most important implication of core stalling is related to the merging of SMBHs. Dynamical friction is the main mechanism responsible for bringing SMBH binaries close enough such that stellar hardening and gravitational wave emission can take over, leading to coalescence \citep[see][for a review]{Merritt2013}. Detecting SMBH mergers in the $10^4-10^6 M_\odot$ mass range, which potentially carry signatures of SMBH seeding processes \citep{Volonteri2009, Ricarte2018}, is one of the key scientific goals of the Laser Interferometer Space Antenna \citep[LISA;][]{AmaroSeoane2017}. Since the majority of BH mergers targeted by LISA are expected to take place within dwarf galaxies \citep{Barausse2020, Volonteri2020, IzquierdoVillalba2023}, core stalling may be particularly relevant. Indeed, \citet{Tamfal2018} show that the orbital decay of intermediate-mass BHs (IMBHs) in dwarf galaxies depends crucially on the inner density power-law slope of their host dark matter halos. They argue that only cuspy halos are favorable for forming a hard IMBH binary, whereas IMBHs will stall at $50-100$ pc in halos with shallower density profiles \citep[see also][]{DeCun2023}. 

The notion that core stalling may inhibit the formation of SMBH binaries raises the exciting possibility of probing the central structure of dark matter halos, and thus also the nature of dark matter, using LISA merger rates. However, this requires that we first develop a better understanding of dynamical friction and of the conditions under which core stalling and dynamical buoyancy operate. Ultimately, any theory for dynamical friction based on Chandrasekhar's formulation is subject to large uncertainties related to the treatment of the Coulomb logarithm \citep[][]{MoVdBWhite, Just2011,Petts2015}, which arise because Chandraskehar's derivation is based on the unrealistic assumption of an infinite homogeneous medium. By properly accounting for the fact that galaxies are inhomogeneous systems, \citet{Tremaine1984} formulated a theory for dynamical friction in spherical systems using linear perturbation theory and showed that dynamical friction is a consequence of the LBK torque (see Section~\ref{sec:background}). \citet{Kaur2018} showed that this gives a natural explanation for core stalling, but their treatment did not explain buoyancy. \citet{Banik2021} relaxed some of the assumptions inherent in the derivation of the LBK torque, and argued that dynamical buoyancy is an outcome of a `memory effect' that arises from the fact that the torque exerted on the perturber depends on its infall history. In a follow-up paper, \citet{Banik2022} presented an alternative explanation for buoyancy: reversal of the torque exerted by a special class of orbits in the restricted three-body framework for dynamical friction. More recently, \citet{Dattathri2025} showed that cores can sometimes be unstable to a rotating dipole mode, and suggested that this may be the origin of dynamical buoyancy.

Although these studies have given important new insight into the dynamics of cored systems, we currently lack a consensus regarding what causes core stalling and/or buoyancy, and under what conditions they operate. In particular, which systems exhibit core stalling and/or buoyancy? Does it depend on the logarithmic gradient of the central density profile, or perhaps on some specific aspect of the kinematics? There are already indications that merely having a density profile for which $\rmd\log\rho/\rmd\log r$ asymptotes to zero at small scale is neither a necessary nor a sufficient condition. For example, \citet{Goerdt2010} used numerical simulations to show that even a system with a moderately cuspy density profile, with $\rho \propto r^{-0.5}$ in the center, can display core stalling, while \citet{Meadows2020} presented a cored system (an isothermal sphere with $\lim_{r\to 0}\rmd\log\rho/\rmd\log r = 0$) that did not reveal any signs of dynamical buoyancy. A fundamental challenge for addressing the necessary conditions for core stalling and/or buoyancy is that the infall of a massive perturber itself perturbs the system in a non-linear manner. This was evident in the simulations of \citet{Goerdt2010}, in which the inspiral of the massive perturber transformed the initial $r^{-0.5}$ density cusp into a core, which then resulted in stalling. This underscores that addressing the conditions for core stalling requires a methodology that can self-consistently account for the evolution of the system subject to the sinking of a massive perturber. While some aspects of this problem can be addressed using quasi-linear theory \citep[][]{Weinberg2001a,Weinberg2001b,Banik2025}, arguably the most straightforward method is to rely on numerical simulations. 

In this paper, we use idealized high-resolution N-body simulations combined with arguments based on kinetic theory to study the motion of massive perturbers in systems with different density profiles. Our main goal is two-fold; to develop a better understanding of the dynamics that give rise to both core stalling and dynamical buoyancy, and to identify the necessary and sufficient conditions for these processes to operate. We will show that it is not just the central gradient in the density profile, but rather the overall shape of the host's distribution function (DF), that ultimately determines the strength of dynamical friction and the emergence of core stalling and buoyancy. Our work follows directly from \citet{Dattathri2025}, who examine the dynamical structure of double power-law density profiles and identify a region of parameter space where the DF has an inflection, resulting in the growth of an unstable dipole mode (the $l=1$ equivalent of the classic radial orbit instability). We extend this work here to show that an unstable dipole mode results in outward motion of massive objects, i.e. dynamical buoyancy. Since the DF is determined by the overall density profile of the system, not just the inner power-law slope, our overall conclusion is that not all cores are equal.

This paper is organized as follows. In Section~\ref{sec:background}, we provide background on the phase-space structure of systems with double power-law density profiles and briefly describe the kinetic theory of dynamical friction. Section~\ref{sec:methodology} describes the N-body simulations that we use in this work, as well as the methodology used to analyze these simulations. Section~\ref{sec:results_param_space} describes the results of our N-body simulations with an infalling perturber, where we show that massive perturbers in systems with very similar density profiles can show markedly different stalling behaviors. In Section~\ref{sec:fE_connection}, we demonstrate that our simulation results can be well explained by studying the phase-space distribution function of the system. Section~\ref{sec:buoyancy} examines the origin of dynamical buoyancy. We put our results in perspective in Section~\ref{sec:discussion} by discussing the underlying dynamics, comparison with other related studies, and the limitations of our model. We discuss the various astrophysical implications of our findings in Section~\ref{sec:implications} and conclude in Section~\ref{sec:conclusions}.

Throughout this paper, we refer to the massive perturber, which we model as a point particle of fixed mass, as a black hole (BH). We emphasize, though, that most of the results will also hold if the massive perturber were any other object (i.e., a satellite galaxy or globular cluster), as long as its tidal stripping is negligible. 

\section{Background}
\label{sec:background}

\subsection{Parameter space of double power law profiles}
\label{sec:density_profiles}

Galaxies and dark matter halos (which we collectively refer to as ``systems'') can be well modeled by a double power law density profile. This profile can be characterized by the general $\abg$ profile \citep{Zhao1996}:
\begin{equation}
\label{eq:dens}
    \rho(r) = \rho_s \, \left( \frac{r}{r_\rms} \right)^{-\gamma} \left[ 1 + \left( \frac{r}{r_\rms} \right)^{\alpha} \right]^{\frac{\gamma-\beta}{\alpha}}  \,.
\end{equation}
Here, $r_\rms$ is the scale radius and $\rho_s$ is a normalization constant. This profile transitions from an outer power-law, $\rho \propto r^{-\beta}$, to an inner power-law, $\rho \propto r^{-\gamma}$, with $\alpha$ controlling the steepness of the transition. We normalize all the $\abg$ profiles such that the total mass $M_{\rm tot}$ and scale radius $r_s$ are equal across all the systems\footnote{Note that this is only possible for systems with a finite total mass, which requires $\beta>3$.}. Several commonly encountered density profiles in astrophysics are examples of the $\abg$ profile, including the NFW \citep{Navarro.etal.1997}, \citet{Hernquist1990}, and \citet{Plummer1911} profiles, which have $\abg=(1,3,1)$, $(1,4,1)$, and $(2,5,0)$ respectively. 

For spherical isotropic systems, the phase-space DF $f(E)$ is given by the \citet{Eddington.16} formula:
\begin{equation}
\label{eq:eddington}
    f(E) = \frac{-1}{\sqrt{8}\pi^2} \frac{\rmd}{\rmd E} \int_0^{E} \frac{\rmd\rho}{\rmd\Phi} \frac{\rmd\Phi}{\sqrt{\Phi-E}}\,.
\end{equation}
Here $\Phi$ is the gravitational potential and $E =\frac{1}{2}v^2 + \Phi$ is the specific energy \citep[see][]{BT2008}. Physical systems must have a positive phase-space density, that is, $f(E)>0$ at all $E$. According to Antonov's stability theorem, systems with $\rmd f/\rmd E <0$ at all $E$ are stable against radial and nonradial perturbations \citep{Antonov1962, Henon1973, BT2008}. 

All three parameters ($\alpha$, $\beta$, and $\gamma$) together determine the overall shape of the DF. However, not all $\abg$ profiles can be supported by a positive, monotonically decreasing DF. \citet{Baes2021} showed that for every set of $(\beta,\gamma)$, there exists a critical value of $\alpha_{\rm unp}$ such that all models with $\alpha>\alpha_{\rm unp}$ are unphysical, i.e. $f(E)<0$ for some range of $E$. Furtherfore, \citet{Dattathri2025} show that there exist a subset of $\abg$ profiles that are physical, but with an inflection point in their DF such that $\rmd f/\rmd E>0$ for a range of $E$. Such systems violate Antonov's stability criterion and are thus not guaranteed to be stable. 

\begin{figure}
    \centering
    \includegraphics[width=\columnwidth]{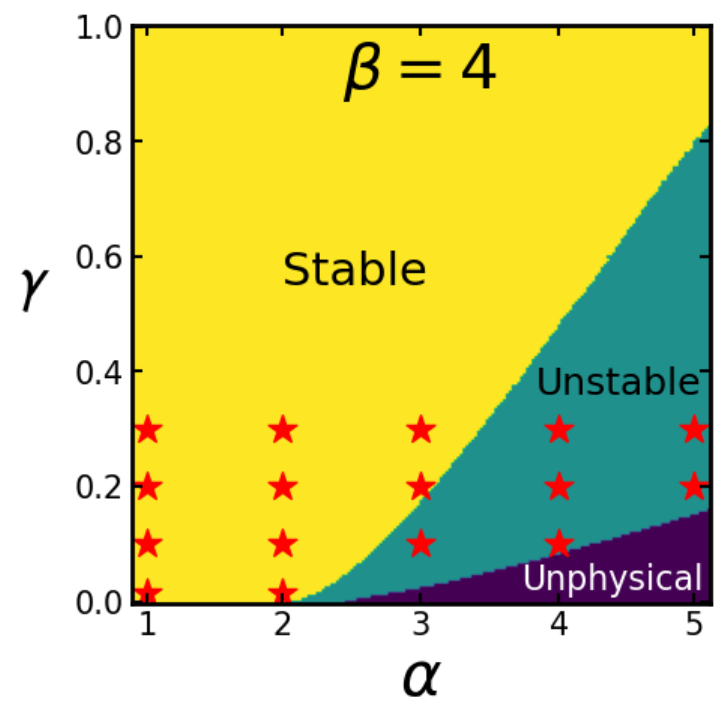}
    \caption{Parameter space of stable, unstable, and unphysical $\abg$ systems in the $\alpha-\gamma$ plane, for fixed $\beta=4$. The red stars indicate the systems that constitute our {\tt Fiducial} suite of simulations.}
    \label{fig:param_space}
\end{figure}

Figure~\ref{fig:param_space} shows, for fixed $\beta=4$, the regions in the $\alpha-\gamma$ plane that correspond to systems that are stable, unstable\footnote{Here defined as systems with an inflection in $f(E)$. Although it has not been formally proved that all such systems are unstable, all the examples studied by \citet{Dattathri2025} were indeed shown to be unstable to a growing dipole mode. Hence we tentatively refer to all such systems as ``unstable''.}, and unphysical. Clearly, unstable systems are characterized by high values of $\alpha$ and low values of $\gamma$, i.e., the density profiles sharply transition from a steep outer power-law slope to a shallow inner slope. Cored profiles (i.e., low $\gamma$) are more easily susceptible to $f(E)$ inflections than cuspy profiles (high $\gamma$). Note that the stable-unstable and unstable-unphysical boundaries depend weakly on the value of $\beta$ (see Figure 2 in \citealt{Dattathri2025}). The red stars indicate the parameters of the density profiles considered in this study. This includes $\alpha=1,2$ for $\gamma=0$, $\alpha=1,2,3,4$ for $\gamma=0.1$, and $\alpha=1,2,3,4,5$ for $\gamma=0.2$ and $\gamma=0.3$. All systems considered here have an outer density slope $\beta = 4$. 
\begin{figure*}
    \centering
    \includegraphics[width=\textwidth]{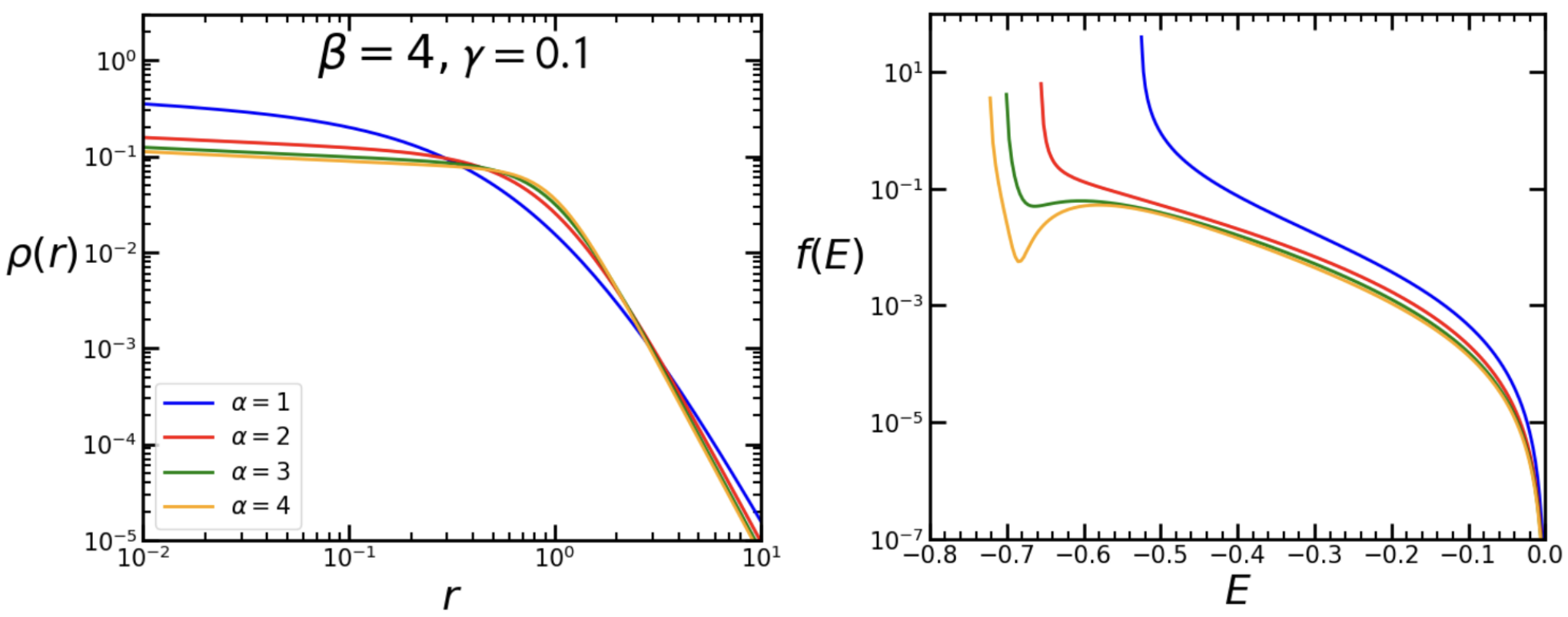}
    \caption{Left: density profiles $\rho(r)$ vs $r$ for four $\abg$ systems, where the total mass is normalized to $1$ across all profiles. We fix $\beta=4$ and $\gamma=0.1$, and vary the value of $\alpha$ between 1 and 4. Right: the isotropic DFs corresponding to these systems. Despite having similar densities, these DFs differ drastically.}
    \label{fig:profiles}
\end{figure*}

Several studies of dynamical friction and the inspiral of massive black holes \citep[BHs, e.g.][]{Dosopoulou2017, Tamfal2018} have focused on varying the logarithmic gradient of the inner density profile (here $\gamma$). However, it is clear from Figure~\ref{fig:param_space} that the DF is also strongly dependent on $\alpha$. This is further illustrated in Figure~\ref{fig:profiles}, the left panel of which shows the density profiles for four $\abg$ profiles that differ only in the value of $\alpha$, as indicated. They all have $\beta=4$ and $\gamma=0.1$ (they correspond to the red stars at $\gamma=0.1$ in Figure~\ref{fig:param_space}), and share the same total mass and scale radius. Although the density profiles of these four systems are very similar, especially those with $\alpha=2$, $3$, and $4$, their DFs, plotted in the right-hand panel of Figure~\ref{fig:profiles}, show drastic differences. In particular, the $\alpha=1$ and $\alpha=2$ profiles have $\rmd f/\rmd E<0$ at all $E$, and therefore are stable. On the other hand, the DFs for the systems with $\alpha=3$ and $\alpha=4$ have an inflection near $E=0.65$, where $\rmd f/\rmd E>0$. As we demonstrate in Section~\ref{sec:buoyancy}, the presence of such an inflection is closely related to dynamical buoyancy. 

The $\abg=(2,4,0.1)$ and $(4,4,0.1)$ systems are widely used throughout the paper. For convenience, we refer to them as the {\tt GradualCore} and {\tt RapidCore} systems, respectively, where ``Gradual'' and ``Rapid'' refer to the transition between the outer and inner power-law slopes.

\subsection{Kinetic theory of dynamical friction}
\label{sec:kinetics}

Before we present the simulation results, we first provide a brief description of the theory for secular evolution, which provides insight into why the DF of the system is the key determinant in the rate of orbital decay. 

Dynamical friction occurs because as a massive perturber orbits a gravitational system, it induces a perturbation, which then back-reacts on the perturber. If the perturbing potential is small, then the response can be analyzed using linear perturbation theory. On the other hand, if the perturbing potential is comparable to the host potential, explicit non-linear effects become important. Below, we discuss these two regimes in detail. 

\subsubsection{Linear perturbation theory and the LBK torque}
Let the DF of the system be $f$ and its Hamiltonian $H$, which obey the collisionless Boltzmann equation:
\begin{equation}
\label{eq:cbe}
    \frac{\partial f}{\partial t} + \left[ f,H \right]=0\,,
\end{equation}
where the square brackets indicate the Poisson bracket. We now perturb these quantities to linear order:
\begin{equation}
\label{eq:pert}
    f = f_0 + f_1 \quad \quad H = H_0 + \Phi_1
\end{equation}
where we have assumed the existence of an equilibrium (not necessarily stable) solution $(f_0,H_0)$. Here, $f_1$ is the total linear response of the system due to the perturbation, and $\Phi_1$ is the total perturbing potential, given by $\Phi_1=\Phi_{\rm BH}+\Phi_{\rm resp}$, where $\Phi_{\rm BH}$ is the potential of the BH, and $\Phi_{\rm resp}$ is the potential due to the system's response. The perturbed DF and response potential are related by the Poisson equation:
\begin{equation}
\label{eq:poisson}
    \nabla^2\Phi_{\rm resp} = 4 \pi G \int f_1 \ d^3 v
\end{equation}
Substituting equation~(\ref{eq:pert}) into equation~(\ref{eq:cbe}), we obtain after linearization:
\begin{equation}
\label{eq:linear}
    \frac{\partial f_1}{\partial t} + \left[f_1,H_0\right]+\left[f_0,\Phi_{\rm BH}\right]+\left[f_0,\Phi_{\rm resp}\right] = 0.
\end{equation}
The above equation describes the evolution of the linear order response of a system to a perturbing potential.

Proceeding further, \citet{Tremaine1984} ignore the response potential $\Phi_{\rm resp}$, and calculate the system's response to the BH alone. This response exerts a torque on the perturber, which in most cases, slows it down, i.e. exerts dynamical friction. For a perturber on a circular orbit, \citet{Tremaine1984} showed that in the time-asymptotic limit ($t \rightarrow \infty$), the torque is given by the LBK torque formula \citep[][]{LyndenBell1972}: 
\begin{equation}
\label{eq:LBK}
    \tau_{\rm LBK} = 16 \pi^4 \sum_{\mathbf{\ell}} l_3 \int \rmd \mathbf{J} \left| \hat{\Phi}_\ell(\mathbf{J}) \right|^2 \mathbf{\ell} \cdot \frac{\partial f}{\partial \mathbf{J}} \delta(\mathbf{\ell} \cdot \mathbf{\Omega} - l_3 \Omega_\rmp).
\end{equation}
Here, $\mathbf{J}=(J_1,J_2,J_3)$ are the three actions, $\mathbf{\Omega} = (\Omega_1,\Omega_2,\Omega_3)$ are the corresponding frequencies, $\mathbf{\ell}=(l_1,l_2,l_3)$ is a vector of integers, $\Omega_\rmp$ is the frequency of the perturber, $\left| \hat{\Phi}_l(\mathbf{J}) \right|$ are the Fourier modes of the perturber potential, $\delta(x)$ is the Dirac delta function, and $f=f(J_1,J_2,J_3)$ is the unperturbed DF. The summation runs from $-\infty$ to $\infty$ for $l_1$ and $l_2$, and from $0$ to $\infty$ for $l_3$. In the case of spherical isotropic systems, the DF is a one-dimensional function of energy, so the LBK torque simplifies to:
\begin{equation}
\label{eq:LBK_isotropic}
    \tau_{\rm LBK} = 16 \pi^4 \Omega_\rmp \sum_{\mathbf{\ell}} l_3^2 \int \rmd \mathbf{J} \left| \hat{\Phi}_\ell(\mathbf{J}) \right|^2 \frac{\rmd f}{\rmd E} \delta(\mathbf{\ell} \cdot \mathbf{\Omega} - l_3 \Omega_\rmp).
\end{equation}
This expression indicates that: (i) angular momentum (and also orbital energy) is exchanged between the perturber and the system exclusively at the resonances that satisfy the commensurability condition $\mathbf{\ell} \cdot \mathbf{\Omega} - l_3 \Omega_\rmp=0$, and (ii) it is the gradient of the DF in action space (in energy space, for isotropic systems) that determines the direction of flux transfer. This is characteristic of secular transport in gravitational N-body systems \citep[][]{Sellwood2014, Hamilton.Fouvry.24}, which underlies a variety of dynamical processes such as, in addition to dynamical friction, the formation of spiral structure, bar formation, and the radial migration of stars \citep[][]{LyndenBell1972, Sellwood.Wilkinson.93, Sellwood2002}.

Note that the LBK torque sums over all resonances. For a stable isotropic system with $\rmd f/\rmd E < 0$  at all $E$, and thus at all resonances, the resulting torque will always be negative. This implies that the LBK torque results in angular momentum loss of the perturber, giving rise to dynamical friction. \citet{Kaur2018} have shown that within the core region of a system, the LBK torque is dominated by the contribution from the corotation resonance (CR) \citep[see also][]{Kaur2022}. This suggests that the sign and magnitude of the LBK torque is governed by the gradient $\rmd f/\rmd E$ at the energy corresponding to CR, which we refer to as $(\nabla f)_{\rm CR}$ hereafter.

While the LBK torque formula gives us powerful insight into the inner workings of dynamical friction, it is based on several assumptions. Firstly, the LBK formula maps the torque onto the {\it unperturbed} DF of the system, and does not account for its evolution. In reality, as the perturber inspirals, it heats up the system, causing the particles to diffuse in energy space (diffusion in the angular momentum can be neglected for a spherical isotropic system). Quasilinear theory dictates that this dynamical friction driven heating modifies the DF via the following diffusion equation \citep[][]{Banik2025}:
\begin{align}
\frac{\partial f}{\partial t} = \frac{\partial}{\partial E} \left(D(\bJ) \frac{\partial f}{\partial E} \right),
\label{diff_eq_E}
\end{align}
where the diffusion coefficient $D(\bJ)$ is given by
\begin{align}
D(\bJ) = \pi\,\Omega^2_\rmP \sum_{\ell} l_3^2\, {\left|\hat{\Phi}_{\ell}\left(\bJ\right)\right|}^2\,\delta\left(\ell\cdot{\bf \Omega} - l_3\Omega_\rmP\right).
\label{D_E}
\end{align}
Quasilinear evolution of the DF and LBK-driven inspiral occur hand in hand and together conserve the total energy and angular momentum, which is evident from the fact that $\tau_{\rm LBK} = -{\left(2\pi\right)}^3 \int \rmd\bJ\, L_z\, (\partial f/\partial t) = -(2\pi)^3 \Omega_\rmP^{-1} \int \rmd\bJ\, E\, (\partial f/\partial t)$. Analogous to the LBK torque, the quasilinear diffusion coefficient is an aggregate of all resonant  contributions. $D(\bJ)$ spikes at each resonant energy, where, given enough time, it generates a plateau in the DF $(\rmd f/\rmd E = 0)$. Since only the co-rotation resonance dominates in the central region of a shallow cusp or a core, we shall see that as the perturber enters the inner region it can create a pronounced plateau in the DF.

Secondly, the LBK torque formula neglects the self-consistent response potential $(\Phi_{\rm resp})$ of the system, which can slo contribute to the system's response to an external perturbation. For example, \citet{Weinberg1989} showed that the system's self-gravity significantly increases the dynamical friction timescale. In addition, self-gravitating systems host a spectrum of point modes which describe the system's fluctuations about its equilibrium \citep{Kalnajs.77, Fridman.Polyachenko.84, Weinberg1994, Palmer1987,BT2008,Petersen2024}. These modes can be oscillating, growing or damped, depending on the initial DF of the system. While damped modes can result in the BH undergoing small-scale wobbles \citep{Weinberg1994, Heggie2020}, more relevant for this paper is if the system possesses an unstable dipole mode due to an inflection in its DF, which as we shall see, can lead to the outward motion of the BH.

Finally, the LBK torque was derived in the asymptotic limit ($t \rightarrow \infty$) of adiabatic growth of the perturber (i.e. the perturber is introduced into the system on a timescale that is long compared to all other relevant timescales). If this assumption is relaxed, as shown in \citet{Weinberg2004} and \citet{Banik2021}, the finite-time response deviates from the LBK prediction by introducing transients. In addition, if the system contains an unstable growing mode, the torques arising from the mode can dominate over the LBK torque. As we shall show in Section~\ref{sec:buoyancy}, this is essentially what is responsible for dynamical buoyancy. 

\subsubsection{Non-linear effects}

The dynamics discussed above are based on linear perturbation theory, which assumes that the perturbing potential $\Phi_1$ is small compared to the host potential. However, as $\Phi_1$ becomes large, non-linear effects such as orbit trapping become important.  The LBK torque is therefore only valid in what \citet{Tremaine1984} refer to as the ``fast regime'', when the perturber is sinking fast enough such that it sweeps through the resonances without any of them building up to non-linear amplitudes. In the ``slow regime'', particles are trapped into librating orbits, which continuously exchange energy and angular momentum with the perturber. If the perturbing potential is small compared to the background potential, the size of the libration zone is small. Moreover, if the perturber itself is stalling, the libration zone itself will remain largely stable and evolve only slowly over time. Since these trapped orbits each have different libration frequencies, they will phase-mix over time. Once the libration zone is fully phase-mixed, the contribution of the trapped orbits to the torque drops to zero \citep{Chiba2022, Chiba2023}. 

Therefore, both in the linear and non-linear regimes (during the stalling phase), the net amount of energy/angular momentum transferred between the perturber and the system is governed by the gradient in the DF\footnote{This is analogous to how a (positive) negative gradient in the DF of a plasma at the group velocity of a Langmuir wave causes (inverse) Landau damping.}. As seen in Figure~\ref{fig:profiles}, systems with very similar density profiles can have drastically different DFs due to a difference in their $\alpha$ values. This warrants a thorough investigation of the inspiral of massive perturbers in different $\abg$ profiles, which is the subject of this paper. 

\section{Methodology}
\label{sec:methodology}

\subsection{Initial conditions}
\label{sec:IC}

We set up spherical isotropic initial conditions for each host system by sampling phase-space coordinates from equations (\ref{eq:dens}) and (\ref{eq:eddington}). We scale the units such that the gravitational constant $G$, the system's total mass (excluding the perturber) $M_{\rm tot}$, and the scale radius $r_s$ are all equal to unity. Throughout the paper, all the results are presented in terms of these model units. 

As detailed in Section~\ref{sec:intro}, the two regimes of interest where cores are especially prominent are the classical dwarfs (which are inferred to live in cored dark matter halos) and massive ellipticals (which have cored surface brightness profiles). Therefore, we use the following scalings to convert our dimensionless units to physical units. For the dwarf halo regime, we adopt the scaling $M_{\rm tot}=3\times 10^9 M_\odot$ and $r_s=1$ kpc, typical of classical dwarfs \citep{Oh2008, Oh2011, Amorisco2013, Read2016}, which gives us a unit time of $T=8.59$ Myr. For the massive elliptical regime, we adopt $M_{\rm tot} = 5 \times 10^{11} M_\odot$ and a characteristic core size $r_s=1$ kpc, typical for brightest cluster galaxies \citep{Lauer2007, Dullo2012, McConnell2012}, yielding a unit time of $T=0.66$ Myr. The BH mass $M_{\rm BH}$ is also expressed in terms of $M_{\rm tot}$. We explore (scaled) BH masses between $10^{-4}$ and $10^{-2}$, typical of BH-galaxy scaling relations \citep{Magorrian1998,Ferrarese2000,Tremaine2002}. These scalings are used in Section~\ref{sec:implications} to interpret the astrophysical implications of our results.
\begin{table*}
\centering

\begin{tabular}{||l|l|c|c|c|l|l||} 
 \hline
 Name & $\alpha$ & $\beta$ & $\gamma$ & $\mbh$ & $N$ & $r_{\rm init}$ \\  
   \hline\hline
    \multirow{1}{8em}{\texttt{Fiducial}} & $1,2$ & $4$ & $0$ & $10^{-3}$ & $10^6$ & 1 \\ 
    & $1,2,3,4$ & $4$ & $0.1$ & $10^{-3}$ & $10^6$ & 1 \\ 
    & $1,2,3,4,5$ & $4$ & $0.2$ & $10^{-3}$ & $10^6$ & 1 \\ 
    & $1,2,3,4,5$ & $4$ & $0.3$ & $10^{-3}$ & $10^6$ & 1 \\ 
    \hline \hline 
  \multirow{1}{4em}{\texttt{MassTest}} & $4$ & $4$ & $0.1$ & $10^{-4}, 3 \times 10^{-4},3 \times 10^{-3}, 10^{-2}$ & $10^6$ & 1 \\ 
   \hline\hline
   \multirow{1}{8em}{\texttt{IsoHighRes}} & $4$ & $4$ & $0.1$ & $0$ & $5 \times 10^6$ & - \\
  \multirow{1}{4em}{\texttt{HighRes}} & $2,4$ & $4$ & $0.1$ & $10^{-3}$ & $5 \times 10^6$ & 1 \\ 
   & $4$ & $4$ & $0.1$ & $10^{-2}$ & $5 \times 10^6$ & 1 \\ 
      \hline \hline 
  \multirow{1}{4em}{\texttt{Buoyancy}} & $1,2,3,4$ & $4$ & $0.1$ & $10^{-3}$ & $10^6$ & 0 \\ 
& $5$ & $4$ & $0.3$ & $0,10^{-4},10^{-3}$ & $ 5\times 10^6$ & 0 \\ 
  \hline \hline
  \multirow{2}{4em}{\texttt{Convergence}} & $4$ & $4$ & $0.1$ & $10^{-3}$ & $10^6$ & 2.5, 5 \\
  & $4$ & $4$ & $0.1$ & $10^{-3}$ & $10^4$, $10^5$ & 1 \\
  & $4$ & $4$ & $0.1$ & $10^{-3}$ & $10^4$, $10^5$, $5 \times 10^6$ & 0 \\
  & $4$ & $4$ & $0.1$ & $10^{-3}$ & $10^6$  & 1 (eccentric) \\ 
 \hline
\end{tabular}
\caption{Summary of the simulations presented in this work. The parameters $\alpha,\beta$, and $\gamma$ control the density profile of the host system (see equation~[\ref{eq:dens}]). The values of the BH mass and its initial radius are given by $\mbh$ and $r_{\rm init}$ respectively. Note that in some cases, there is no BH. The value $N$ indicates the total number of field particles in the simulation (excluding the BH, if it exists).}
\label{table:sims}
\end{table*}

In most of our simulations, we instantaneously introduce a point particle of mass $\mbh$, representing the BH, into the system. The BH is initialized either on a circular orbit with radius $r_{\rm init}$, or at rest at the center of the system. The only exception is a small suite of simulations discussed in Appendix~\ref{app:eccentric} for which we initialize the BH (instantaneously) on an eccentric orbit and which results in stalling behavior that is qualitatively similar to that in the case of circular orbits.

Note that we always introduce the BH instantaneously into the system. In principle, as mentioned in Section~\ref{sec:background}, this can introduce transients that cause the net torque on the BH to deviate from the LBK prediction \citep{Weinberg2004, Banik2021}. We have run various experiments to assure that these transients do not significantly impact any of our results. An example is shown in Appendix~\ref{sec:appendix}, where we show that the stalling experienced by an infalling BH is independent of the initial radius $r_{\rm init}$, at which the BH is instantaneously introduced.

\subsection{N-body simulations}
\label{sec:Nbody}

All of our simulations are run using the publicly available N-body code {\sc GyrfalcOn} \citep{Dehnen2002}. In most simulations, the host system is sampled with $N=10^6$ equal mass particles. We also rerun a few simulations at higher resolution with $N=5 \times 10^6$ particles, which are used for a more detailed examination of the phase-space evolution in Section~\ref{sec:fE_connection}. Appendix~\ref{sec:appendix} presents convergence tests that show that in general $10^6$ particles suffice to resolve the intricate dynamics related to core stalling and buoyancy.

All simulations are run using hierarchical time stepping, where the time step of each particle depends on its softening length and instantaneous acceleration, $a_i$, according to $\Delta t \sim 0.05 \sqrt{\epsilon_i/|a_i|}$. With this criterion, the circular orbital time around the BH at a separation $\epsilon$ is resolved with $\sim 100$ timesteps. We assign softening lengths to the particles based on the \citet{Dehnen2001} criterion:
\begin{equation}\label{softening}
 \epsilon = 0.017 \left( \frac{N}{10^5} \right)^{-0.23} .
\end{equation}
The BH is assigned a softening length equal to half of that of the particles, i.e. $\epsilon_{\rm BH}= 0.5 \epsilon$. For our high-resolution simulations, we have $\epsilon=7\times 10^{-3}$ and $\epsilon_{\rm BH}=3.5 \times 10^{-3}$. 

\subsection{List of simulations}
\label{sec:lists}

Table~\ref{table:sims} lists the various simulations discussed in this paper. These vary in the density profile of the host halo, in particular, the inner cusp slope $\gamma$ and the parameter $\alpha$ that controls how rapidly the outer profile transitions to the inner profile, the mass and initial radius of the BH, and the mass resolution of the simulation. The details of the various simulation suites listed are expanded upon when we introduce them in the subsequent sections.
\begin{figure*}
    \centering
    \includegraphics[width=\textwidth]{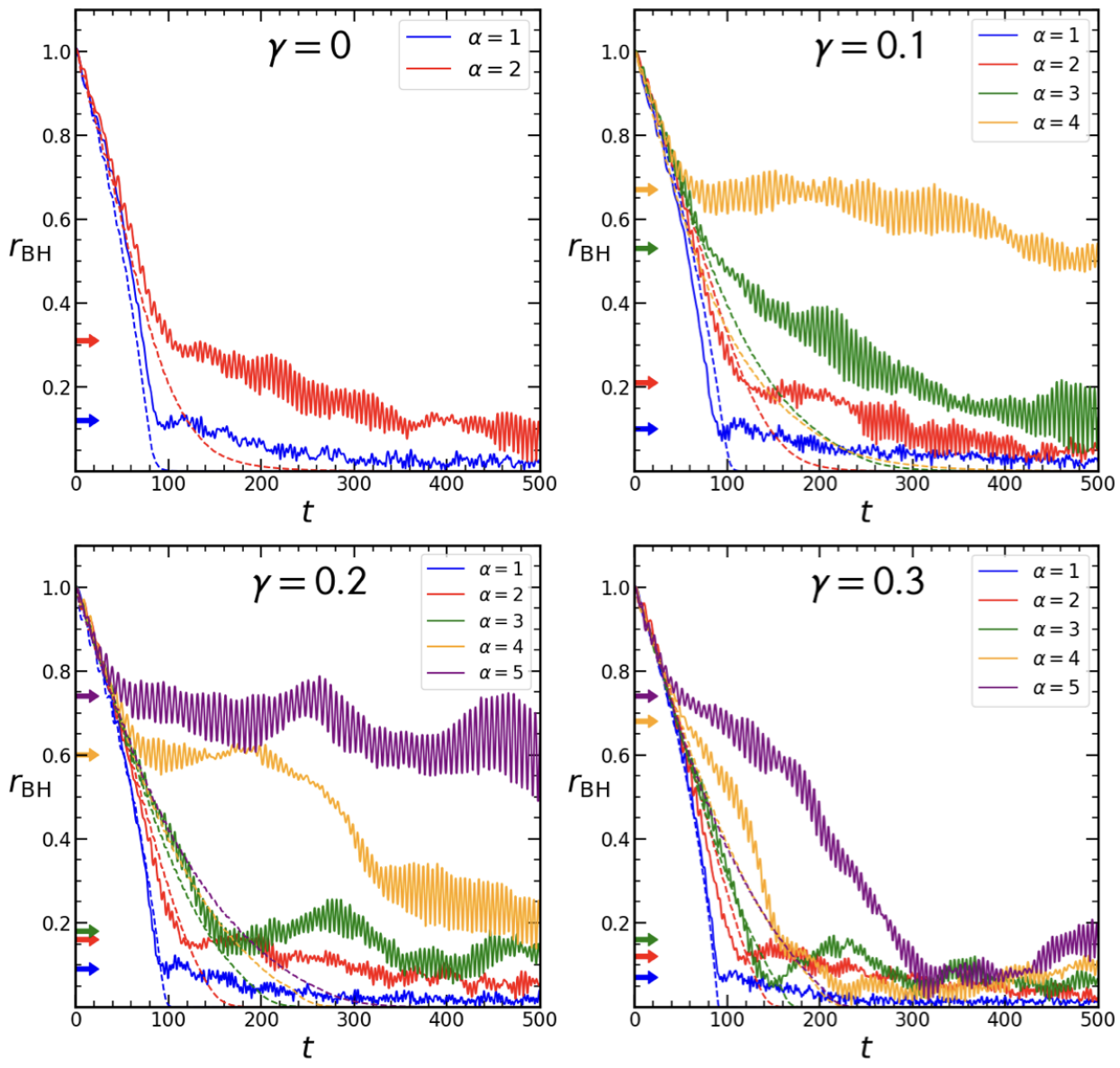}
    \caption{Inspiral of a BH of mass $\mbh=10^{-3}$ in systems with different density profiles. We fix $\beta=4$ for all cases, and each panel corresponds to a different value of $\gamma$. The dashed lines denote the prediction from the Chandrasekhar prescription (equation~[\ref{eq:c43}]), and the horizontal arrows mark $r_{\rm stall}$, where the trajectory deviates from the prediction. Note how the different $\alpha$ values result in drastically different trajectories.}
    \label{fig:inspiral}
\end{figure*}
\subsection{Calculating the instantaneous distribution function}
\label{sec:instant}

An important goal of this study is to analyze the evolution of the system's DF in response to the inspiraling BH. We compute the DF using the particles' instantaneous energy $E$ and angular momentum $L$ using the method of \citet{Dattathri2025}. We assume that the system remains spherical throughout its evolution, so that the DF is two-dimensional in $E$ and $L$, i.e. $f=f(E,L)$. The DF is calculated from the particle distribution $N(E,L)$ as:
\begin{equation}\label{eq:dNdEdL}
   \frac{\rmd^2 N(E,L)}{\rmd E\,\rmd L} = f(E,L) \, g(E,L)\,,
\end{equation}
where $f(E,L)$ is the DF, and 
\begin{equation}\label{eq:dens_states}
    g(E,L) = 16 \pi^2 L \int_{r_\rmp}^{r_\rma} \frac{\rmd r}{\sqrt{2(E-\Phi(r))-(L/r)^2}} \ ,
\end{equation}
is the density of states \citep{BT2008}. Here $r_\rmp$ and $r_\rma$ are the pericenter and apocenter distances for an orbit with energy $E$ and angular momentum $L$, which, in a spherical potential, are the roots of
\begin{equation}
 \frac{1}{r^2} + \frac{2(\Phi(r)-E)}{L^2} = 0\,.
\end{equation}

Note that the infall of the BH can modify the potential, thus changing $g(E,L)$. Therefore, at each time, we construct the spherically averaged potential $\Phi(r)$, and then use equation~(\ref{eq:dens_states}) to self-consistently obtain $g(E,L)$. This accounts for the changes in the density and potential of the system as the BH sinks in, albeit under the assumption that the system remains spherical.

The system is set up with isotropic initial conditions, such that the DF is one-dimensional in energy, i.e., $f=f(E)$. As the BH sinks in, it perturbs the system and its DF, as dynamical friction transfers energy and angular momentum from the BH to the field particles. However, in practice, we find that even at late times, the DF remains mostly isotropic and develops only a very weak dependence on $L$. Therefore, we mainly focus on the one-dimensional $f(E)$, which we compute by marginalizing $f(E,L)$ over $L$:
\begin{align}\label{eq:marginalization}
f(E) = \dfrac{\int_0^{L_\rmc\left(E\right)} \rmd L\, f(E,L) g(E,L) }{\int_0^{L_\rmc\left(E\right)} \rmd L\, g(E,L)}\,,
\end{align}
with $L_\rmc(E)$ the angular momentum of a circular orbit of energy $E$. For isotropic systems, the $f(E)$ thus obtained is identical to that inferred using the Eddington inversion (equation~[\ref{eq:eddington}]).

\section{Results: stalling in different systems}
\label{sec:results_param_space}

In this section, we begin our exploration of dynamical friction by studying the inspiral rates of massive BHs in the $\abg$ systems indicated by the red stars in Figure~\ref{fig:param_space}. All these simulations are run with $N=10^6$ particles and the BH is introduced instantaneously on a circular orbit at $r_{\rm init}=1$. This section, as well as Section~\ref{sec:fE_connection} and \ref{sec:buoyancy}, describes the simulation results; the underlying dynamics are discussed in Section~\ref{sec:discussion}.
\begin{figure*}
    \centering
    \includegraphics[width=\textwidth]{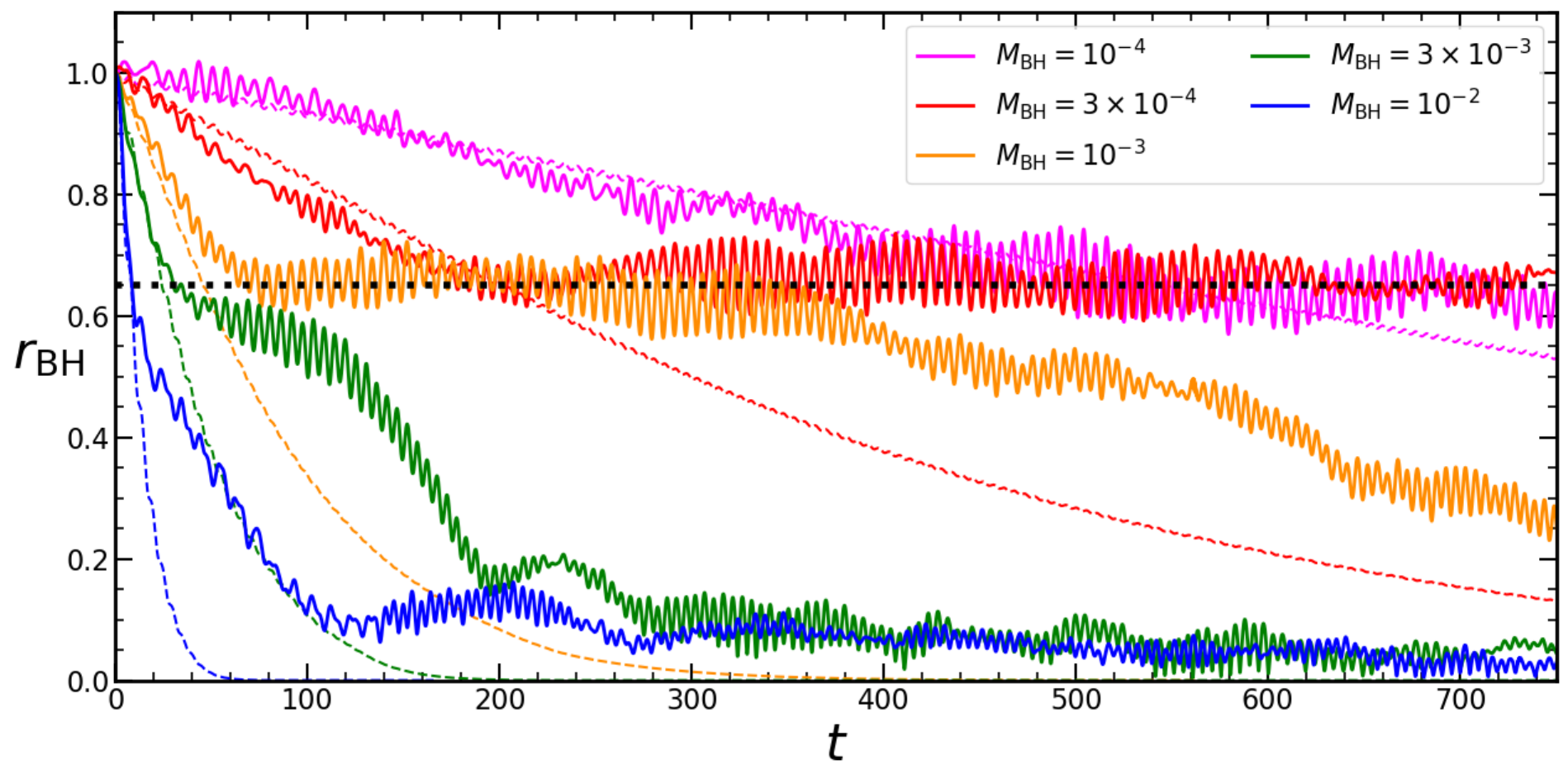}
    \caption{Inspiral of BHs of varying masses, ranging from $\mbh=10^{-4}$ to $10^{-2}$, in the {\tt RapidCore} system. The solid lines show the N-body simulation results, and the dashed lines show the corresponding predictions from the Chandrasekhar prescription (equation~[\ref{eq:c43}]) We see that regardless of mass, all BHs stall at $r_{\rm BH} \approx 0.65$. However, the more massive BHs are able to break through the stalling phase earlier, and sink in (slower than predictions from equation~[\ref{eq:c43}]) down to $r_{\rm BH}\approx 0.1$. The lightest BHs stall perfectly at $r_{\rm BH} \approx 0.6$ and never sink further. }
    \label{fig:pert_mass}
\end{figure*}
    
\subsection{Orbital decay and stalling}
\label{sec:decay}

We start with the \texttt{Fiducial} simulations (see Table~\ref{table:sims}), in which we fix the BH mass at $\mbh=10^{-3}$, and only vary the density profile of the host system. 

Figure~\ref{fig:inspiral} shows the BH's inspiral in the various systems. Different panels correspond to different values of the inner slope $\gamma$, while different colored curves in each panel correspond to different values of $\alpha$, as indicated. The solid lines indicate the distance of the BH from the system's center of mass\footnote{The center of mass is calculated using only the 90\% most bound particles. We also experimented with using the 10\%, 50\% and 95\% most bound particles, all of which yields results that are indistinguishable.} in the simulations, and the dashed lines indicate predictions from the Chandrasekhar prescription (equation~[\ref{eq:c43}]). For the latter, as is commonly done, we assume that the velocity distribution of stars at a particular radius follows a Maxwellian distribution \citep{BT2008}, and we use $\ln \Lambda=5$, which is tuned to match the BH trajectory at early times. 

In all cases, equation~(\ref{eq:c43}) accurately predicts the inspiral at early times. At a certain point, the trajectory deviates from expectation, after which the BH either stalls completely with no further infall or continues to sink but with an inspiral rate that is significantly slower than the Chandrasekhar prediction. In what follows, we refer to both deviations as ``stalling'', although it is clear that in many cases, the BH does not truly stall with zero orbital decay but rather shows reduced decay. The stalling radius for each case is marked with an arrow on the left-hand side.

The BH trajectory has a strong dependence on $\alpha$. For a fixed $\gamma$, as the value of $\alpha$ increases, the stalling occurs at a larger radius. For example, in the case of $\gamma=0.1$, the stalling radius increases from $r_{\rm stall} \approx 0.1$ for $\alpha=1$, to $r_{\rm stall}\approx 0.65$ for $\alpha=4$. Recall that for fixed values of $\gamma$ and $\beta$, the density profiles of different $\abg$ models are roughly similar. However, their distribution functions (DFs) differ significantly (see, e.g., Figure~\ref{fig:profiles}). Therefore, the pronounced variation in the BH trajectories within each panel of Figure~\ref{fig:inspiral} directly reflects the sensitivity of the inspiral rate to the DF. We investigate this further in Section~\ref{sec:fE_connection}.

In most cases, $r_{\rm stall}$ decreases with increasing $\gamma$. In addition, even after the trajectory deviates from the predictions based on equation~(\ref{eq:c43}), the continued inspiral is faster for higher $\gamma$. For example, note the stark difference in trajectories for $\alpha=4$ between the $\gamma=0.1$  and $\gamma=0.3$ cases.  

Finally, we note that while the results shown in Figure~\ref{fig:inspiral} all correspond to circular orbits, we also ran several simulations in which the BH is initially placed on an eccentric orbit. As shown in Appendix~\ref{sec:appendix}, in this case, the orbit of the BH largely maintains its eccentricity as it sinks in and stalls when its {\it average} radius becomes comparable to $r_{\rm stall}$ (see Figure~\ref{fig:convergence}). Hence, we tentatively conclude that our results are qualitatively similar in the case of eccentric orbits. 

\subsection{Dependence on BH mass}
\label{sec:pert_mass}

Next, using the \texttt{MassTest} suite of simulations (see Table~\ref{table:sims}), we investigate how orbital decay depends on the mass of the BH. We vary $\mbh$ between $10^{-4}$ and $10^{-2}$ (representative of typical BH-galaxy scaling relations), keeping the host density profile fixed to that of the {\tt RapidCore} system which has $\abg=(4,4,0.1)$. All other parameters (numerical and physical) are the same as for the {\tt Fiducial} simulations. 

Figure~\ref{fig:pert_mass} shows the various trajectories, with the dashed lines showing the corresponding Chandrasekhar predictions based on equation~(\ref{eq:c43}). Across two orders of magnitude in $\mbh$, the BH trajectory always starts out in close agreement with the Chandrasekhar prediction. However, once the BH reaches the stalling radius $r_{\rm stall} \approx 0.65$, indicated by the black dotted line, it experiences stalling (i.e., its trajectory deviates from the prediction). It is clear, though, that with increasing BH mass, stalling is becoming more and more imperfect. More massive BHs spend less time at $r_{\rm BH}\approx r_{\rm stall}$ and are able to quickly ``break through'' the $r_{\rm stall}$ barrier. For example, in the $\mbh=10^{-4}$ case, the BH remains at $r_{\rm stall}$ for a very long time without any further inspiral. In contrast, the $\mbh=10^{-2}$ trajectory barely ``stalls'' at all, and instead continues its inspiral albeit at a slower rate than the prediction from equation~(\ref{eq:c43}). Interestingly, at late times, the $\mbh=3 \times 10^{-3}$ and $10^{-2}$ BHs stall again at $r_{\rm BH} \approx 0.05$. This secondary stalling radius is significantly larger than the expected oscillation amplitude due to the internal modes of the system \citep{Lau2021}. The origin of this is unclear, but we note that at such small radii, the enclosed mass $M_{\rm enc}$ is comparable to $\mbh$. Hence, the definition of ``center'' becomes somewhat ambiguous, with the central core and BH rotating about each other's common center of mass. In addition, recent work by \citet{DiCintio2025} show that discreteness-induced effects can also lead to stalling at such small radii.

It is notable that the stalling radius is largely independent of $\mbh$, as seen in Figure~\ref{fig:pert_mass}. This goes against predictions from some analytic models based on linear theory \citep{Kaur2018, Banik2021} and semi-analytic models \citep{Petts2015, Petts2016}, which predict that more massive perturbers should stall further out (see Section~\ref{sec:comparison} for a more detailed discussion). 
 
\section{Relation to distribution function}
\label{sec:fE_connection}

The results presented in the previous section demonstrate that the inner slope of the density profile is \textit{not} the most important factor determining the strength of dynamical friction. The dependence on the value of $\alpha$ is particularly striking, as is evident from Figure~\ref{fig:inspiral}. For fixed outer and inner power-law slopes ($\beta$ and $\gamma$), the density profiles are remarkably similar (see Figure~\ref{fig:profiles}), but the extent to which the BHs experience stalling are very different.  

It is clear from Figure~\ref{fig:profiles} that the DF of an $\abg$ system strongly depends on $\alpha$. Specifically, with increasing $\alpha$, the DF \textit{flattens} at high binding energy and eventually develops an inflection where $\rmd f/\rmd E>0$. As discussed in Section~\ref{sec:background}, within the core region, secular evolution theory predicts that the magnitude and direction of the torque depends on the gradient of the DF at the corotation resonance. By definition, a BH orbiting on a circular orbit is always located at corotation. In addition, in a spherical system, the azimuthal frequency $\Omega_\phi$ at fixed energy is nearly independent of angular momentum. This suggests that the direction of the torque it experiences is determined by the gradients of the DF with respect to energy at its own orbital energy $E_{\rm BH}$. In other words,  $(\nabla f)_{\rm CR} = (\rmd f/\rmd E)\big\rvert_{E_{\rm BH}}$. Below, we test this conjecture in detail. 

Accurately measuring the phase-space distribution, especially in the central region, requires high resolution. Therefore, for most of this section, we use the \texttt{HighRes} and \texttt{IsoHighRes} simulations, in which we boost the resolution to $5 \times 10^6$ particles. In Appendix~\ref{sec:appendix} we show that the BH trajectories are nearly indistinguishable between the $N=10^6$ and $N=5\times 10^6$ simulations. 

\subsection{Evolution in isolation}
\label{sec:isoevol}

Before we examine the evolution of the DF in response to the inspiraling BH, we first address an important point. Spherical isotropic systems that have $\rmd f/\rmd E<0$ for all $E$ are stable to radial and non-radial modes \citep{Antonov1962,Doremus1971}. The $\abg$ systems are stable if they are located within the yellow shaded region in Figure~\ref{fig:param_space}. When evolved in isolation, they remain perfectly spherical and isotropic, and their DFs show no evolution apart from some small fluctuations due to discreteness effects. However, systems that have an inflection in their DFs, where $\rmd f/\rmd E >0$, violate Antonov's stability criterion, and all such systems studied by \citet{Dattathri2025} were found to be unstable to a growing dipole mode (see Footnote 3). This dipole instability is the $l=1$ analogue of the radial orbit instability \citep{Barnes1986}, which results in the formation of a bar-like structure (see \citealt{Bortolas2022} for a detailed study on BH inspiral in the presence of bars). As we shortly show in Section~\ref{sec:buoyancy}, an unstable dipole mode results in dynamical buoyancy.

The amplitude of the dipole mode initially grows exponentially with time, in accordance with linear response theory \citep[e.g.][]{Kalnajs.77, Weinberg1991}. Once the mode goes non-linear, it traps particles into librating orbits, and this induces permanent changes in the DF, eroding the initial bump. Saturation of the dipole mode occurs when the DF attains a quasi-equilibrium state, after which the amplitude remains constant. In the system examined in detail by \citet{Dattathri2025}, which corresponds to an $\abg=(6,5,0.5)$ profile, upon saturation the initial inflection in the DF has been mostly erased, although not completely.

\begin{figure}
    \centering
    \includegraphics[width=\columnwidth]{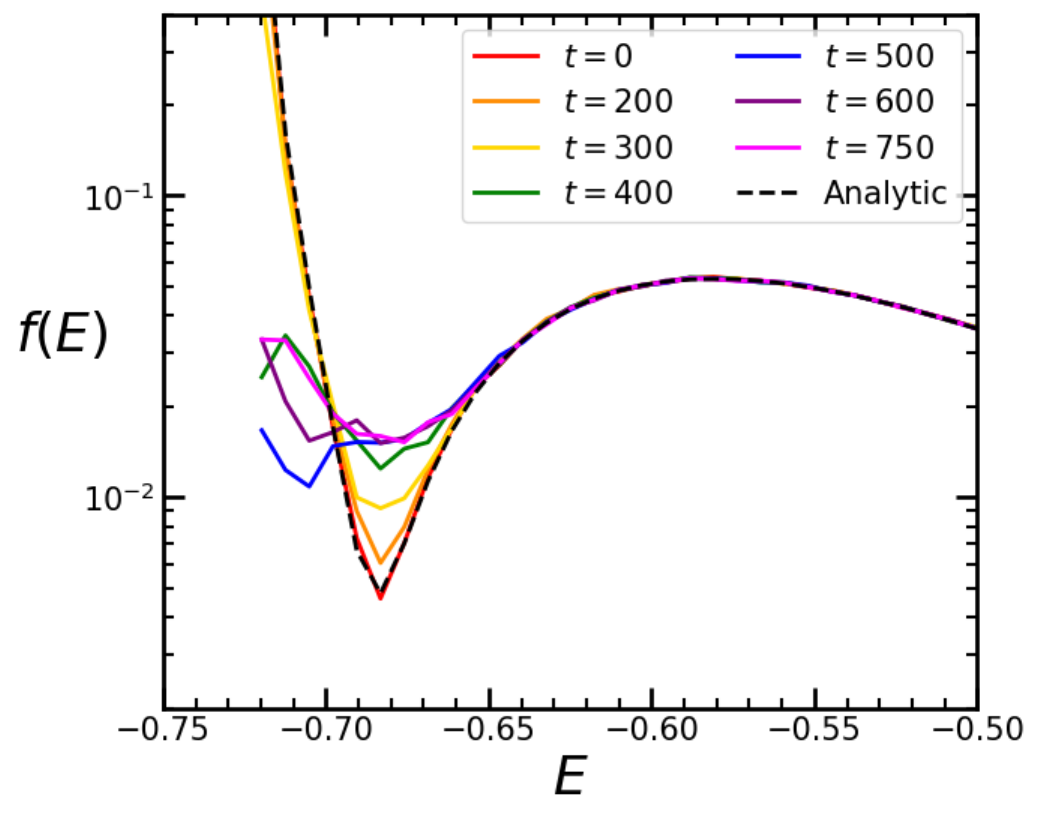}
    \caption{Evolution of the {\tt RapidCore} system's DF in isolation. As the system evolves, the inflection feature in the DF gets backfilled, due to the growth of an unstable dipole mode. However, this does not completely erase the inflection. The DF stabilizes at $T \approx 500$, after which there is no further evolution.}
    \label{fig:fE_nobh}
\end{figure}

For systems with inflections in their DFs, which includes the {\tt RapidCore} system, it is crucial to first understand their evolution in isolation, that is, without a perturber. This provides a baseline for quantifying the relative contributions of secular evolution in the DF driven by the dipole mode and that due to the infalling BH. Hence, we first analyze the \texttt{IsoHighRes} simulation, in which the {\tt RapidCore} system is evolved in isolation. Figure~\ref{fig:fE_nobh} shows the evolution of its DF $f(E)$ over time, as calculated using equation~(\ref{eq:marginalization}). The black line shows the initial $f(E)$ calculated analytically using Eddington inversion (equation~[\ref{eq:eddington}]), which has a pronounced inflection at $E \approx -0.6$. Over time, the inflection in the DF is `backfilled' due to the growth and saturation of an unstable dipole mode, similar to what was observed in \citet{Dattathri2025}\footnote{\citet{Dattathri2025} calculate the amplitude of the dipole mode using EXP \citep{Petersen2025}, which uses biorthogonal basis functions to measure the gravitational power in each harmonic. However, this method does not yield a detectable dipole mode in this case, presumably because the mode's amplitude is too low to distinguish it from the background noise. Using linear response theory and spectral analysis, we have verified that an unstable dipole mode does exist, with a pattern speed of $\Omega_\rmp=0.59$, and that it grows and saturates around $t\approx 500$.}. The mode grows until $t \approx 500$, after which it saturates and there is no further evolution in the DF. 

Interestingly, the backfill of $f(E)$ does not run to completion in that the late-time DF still has a local minimum, and therefore an interval in energy where $\rmd f/\rmd E>0$. What determines the late-time ``equilibrium'' DF of an unstable system as the instability saturates is left for future study, but we suspect that it depends sensitively on the detailed phase-space structure and the topologies of the various resonances. 

\subsection{Relation between inspiral rate and local $f(E)$ gradient}
\label{sec:relation}

\begin{figure*}
    \centering
    \includegraphics[width=2\columnwidth]{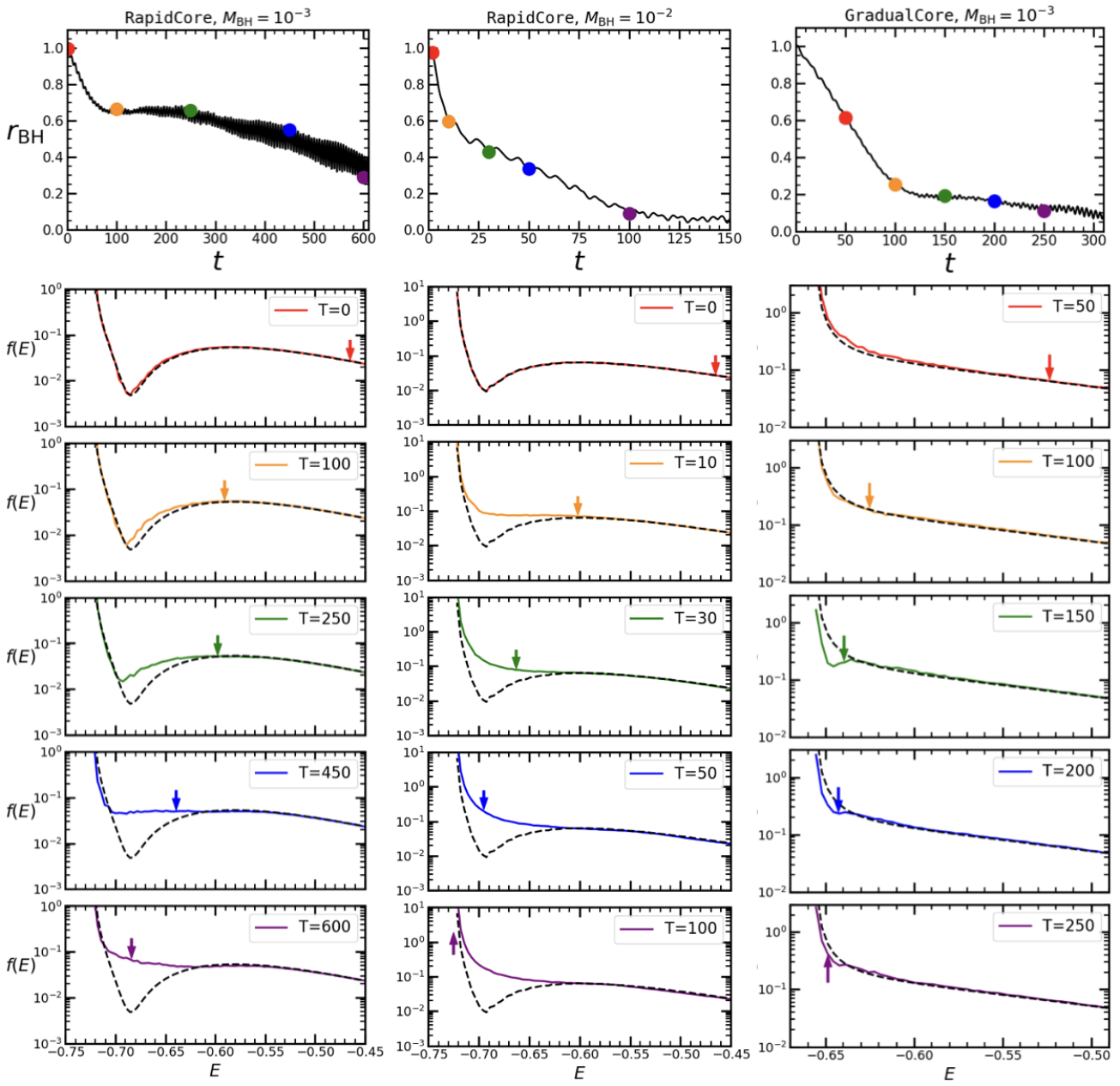}
    \caption{Relationship between BH inspiral and the evolution in the DF, $f(E)$. In each column, the top panels show the BH trajectory, $r_{\rm BH}$ vs $t$. The subsequent panels in each column show the system's DF at different points in time, with the arrow indicating the BH's energy and the black dashed line showing the initial, unperturbed DF. Left panels correspond to the {\tt RapidCore} system with $\mbh=10^{-3}$, middle panels correspond to the {\tt RapidCore} system with $\mbh=10^{-2}$, and right panels correspond to the {\tt GradualCore} system with $\mbh=10^{-3}$. See text for discussion.}
    \label{fig:dist_func_plots}
\end{figure*}

We now analyze the {\tt HighRes} simulation of the {\tt RapidCore} system with $\mbh=10^{-3}$, to study the relation between $\gradf$ and the BH inspiral rate as well as the evolution of the DF of the system over time. The top left panel of Figure~\ref{fig:dist_func_plots} shows the radial trajectory of the BH, $r_{\rm BH}(t)$. The colored curves in the panels below show the DF of the system at various points during the inspiral, as indicated by points of matching color in the top panel. In each panel, the arrow indicates the instantaneous orbital energy of the BH, and thus the energy at corotation.

At early times $(t\lesssim100)$, the BH sinks rapidly through the system. During this phase, $\gradf = (\rmd f/\rmd E)|_{E_{\rm BH}} <0$, and the LBK torque (equation~[\ref{eq:LBK_isotropic}]) indeed predicts a negative torque (i.e. friction) on the perturber. Note that during this period, the inspiral closely matches the \citetalias{C43} prediction based on equation~(\ref{eq:c43}). At $t\approx 100$, the BH reaches a plateau in $f(E)$, where $\gradf \approx 0$. In agreement with our conjecture, the LBK torque now vanishes, causing the BH to stall its inspiral at $r_{\rm BH} \approx 0.65$. 

During the stalling phase, as seen in Figure~\ref{fig:fE_nobh} the DF of the system continues to undergo significant evolution. In particular, the dip/inflection in $f(E)$, which is initially present between $E\approx -0.7$ and $-0.6$, is eroded over time, akin to what occurs in isolation. However, unlike in the isolated case, the inflection/bump is completely eroded by $T \sim 500$. Importantly, once the inflection in the DF has been erased, and the DF has re-established a negative gradient at the orbital energy of the perturber, the LBK torque becomes negative again thereby reinitiating dynamical friction that causes the BH to continue its inward journey.

The fact that in the presence of the BH the host system shows a more pronounced evolution in its DF than in isolation suggests that this elevated evolution is due to the perturber, either by directly modifying the actions of the particles with which it interacts, or by influencing the evolution and/or saturation of the dipole mode that is present as a consequence of the original inflection. To gain some additional insight, we have run tests in which we instantaneously remove the BH from the simulation when it reaches the stalling radius, after which we continue to evolve the system in isolation. Once the BH is removed, we notice no subsequent evolution in the DF, indicating that the strong evolution in the DF evident in Figure~\ref{fig:dist_func_plots} is due to the BH modifying the DF while it is stalling. As discussed in Section~\ref{ssec:nonlinear}, this continued evolution of the DF is mainly due to orbit trapping of particles by the BH. 

\subsection{Dependence on perturber mass}
\label{sec:mpdep}

The above concepts regarding refilling of the inflection in the DF due to the perturber can be used to understand why more massive perturbers are quickly able to ``break through'' the $r_{\rm stall}$ barrier (Figure~\ref{fig:pert_mass}). The panels in the middle row of Figure~\ref{fig:dist_func_plots} show the BH trajectory (top panel) and evolution of the DF (bottom panels) of the {\tt HighRes} simulation of the {\tt RapidCore} system with $\mbh=10^{-2}$ (i.e., an order of magnitude more massive than in the panels in the left-hand column). Due to its larger mass, the BH quickly sinks from $r_{\rm BH}=1$ to $r_{\rm BH}=r_{\rm stall}\approx 0.65$ within $t \approx 10$. This is the point in time at which the trajectory starts to deviates from the  \citetalias{C43} prediction (see Figure~\ref{fig:pert_mass}). Note, though, that unlike in the case with $\mbh=10^{-3}$, by the time the BH has reached the stalling radius the inflection in the DF has already completely backfilled. It is clear that this rapid evolution in the DF is driven by the inspiral of the BH itself. Interestingly, since the gradient in the DF at the orbital energy of the perturber is now always negative, the BH always experiences a negative torque, and thus continues to sink inward, albeit at a rate slower than equation~(\ref{eq:c43}) (see Figure~\ref{fig:pert_mass}).

\subsection{Dynamic creation of an inflection point}
\label{sec:inflectioncreation}

It is clear from the previous subsections that the inspiral rate of the BH is closely related to the local gradient in the DF, $\gradf$. In addition, the inspiral itself can induce permanent changes in the DF, which in turn influence the subsequent radial trajectory of the BH. In particular, we have seen how an inspiraling BH can backfill a dip in the DF, thereby preventing perfect stalling for prolonged time. We now demonstrate that an inspiraling BH can {\it create} an inflection in the DF, even when none is present initially, thus inducing its own subsequent stalling. This dynamic creation of a DF inflection also leads to the growth of a dipole mode in a system which is otherwise stable against dipole instabilities.

The right-hand panels of Figure~\ref{fig:dist_func_plots} show the results of our \texttt{HighRes} simulation of the {\tt GradualCore} system, for which $\abg=(2,4,0.1)$, with a BH of mass $\mbh=10^{-3}$. The unperturbed initial DF of this system does not have any inflection and has $\rmd f/\rmd E<0$ at all $E$. Hence, the system is stable, and we might naively expect that an infalling BH in this system does not experience any stalling. However, as is apparent from the top right panel of  Figure~\ref{fig:dist_func_plots}, this is not the case, as the BH clearly experiences stalling at  $r_{\rm stall} \approx 0.21$. The bottom-right panels of Figure~\ref{fig:dist_func_plots} show the evolution of the DF of the {\tt GradualCore} system over time. During the early stages ($t \lta 100$), the inspiral of the BH is similar to that of the {\tt RapidCore} system: the BH quickly sinks in until it reaches $E_{\rm BH} \approx -0.6$. Between $t=100$ and $t=150$ the BH \textit{creates} an inflection in the DF close to its own orbital energy, triggering a dipole mode. As a consequence,  $\gradf$ becomes sufficiently small so that the BH starts to experience stalling. This newly-created inflection in the DF does not remain permanently, but rather backfills over time. Consequently, by $t \sim 200$ the local gradient in the DF becomes negative again, allowing the BH to continue to spiral inward albeit rather slowly. 

The dynamic creation of inflections in the DFs of systems that are initially stable (and thus have $\rmd f/\rmd E < 0$ at all $E$) might explain why \citet{Goerdt2010} observed stalling even in systems that initially have a density profile with an inner slope as steep as $\gamma = 0.75$. In particular, they find that in such systems the dynamical heating induced by the infalling perturber can transform the central density cusp into a core, after which the perturber stalls. Based on the results presented here, we suspect that this core formation likely involves the creation of an inflection of plateau in the DFs, which then would explain the ensuing stalling. Indeed, in a follow-up paper \citep[][]{vdBosch.Dattathri.25} we show an example of a case similar to that presented here, in which the creation of an inflection in the DF coincides with a rapid tidal disruption of the central region.

\section{Origin of dynamical buoyancy}
\label{sec:buoyancy}

According to the LBK formula, if $\gradf >0$, the corotation resonance will contribute a net positive (enhancing) torque on a perturber, resulting in an outward movement of a massive perturber, i.e. buoyancy. Hence, it is tempting to link buoyancy to the presence of an inflection in the DF. In our four $\abg$ models with $\beta=4$ and $\gamma=0.1$, the $\alpha=1$ and $\alpha=2$ models have $\rmd f/\rmd E<0$ everywhere, whereas the $\alpha=3$ and $\alpha=4$ models have inflection points where $\rmd f/\rmd E>0$ (see Figure~\ref{fig:profiles}). This suggests that  the latter two systems might manifest buoyancy. However, several complications have to be accounted for. First of all, in both systems the dip in the DF is at $E \approx 0.65$. Hence, a BH that is placed near the center of these systems would still have a {\it local} gradient in the DF that is negative. More importantly, as shown by \citet{Dattathri2025}, systems with $\rmd f/\rmd E>0$ are unstable to a dipole mode, which exerts a torque on the central cusp, dislodging it, and setting in into motion. If the center also contains a massive BH, it is to be expected that this torque will also cause the BH to move outward. Hence, buoyancy may well be a direct consequence of the dipole instability, rather than a manifestation of a positive LBK torque.

\begin{figure*}
    \centering
    \includegraphics[width=\textwidth]{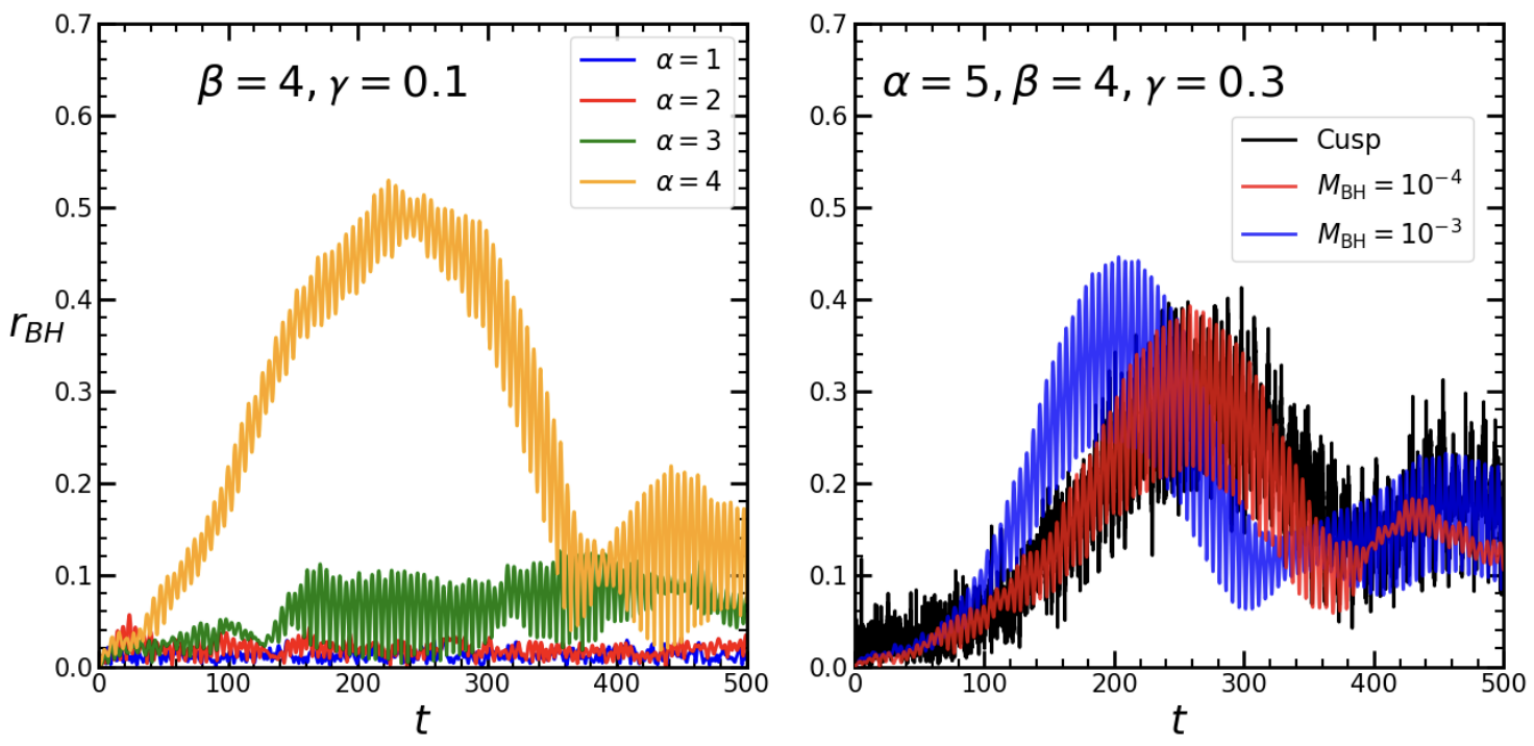}
    \caption{{\it Left:} motion of a BH of mass $\mbh =10^{-3}$ that is initially placed at the center of different $(\alpha,4,0.1)$ systems, with the various curves corresponding to different $\alpha$ values. In the systems with no DF inflections $(\alpha=1,2)$, the BH shows only Brownian motion. On the other hand, in the $\alpha=3,4$ systems, which have an inflection in their DFs, the BH shows a clear outward motion away from the center. The maximum offset is greater in the $\alpha=4$ case, which has a greater dip in the DF (see Figure~\ref{fig:profiles}). {\it Right:} trajectories of the cusp (black curve) and BHs of mass $\mbh=10^{-4}$ and $10^{-3}$ (red and blue curves, respectively) in the $\abg=(5,4,0.3)$ system. Due to an inflection in the DF, the system develops a growing dipole mode, which is responsible for dislodging the cusp and the BHs from the center, setting them in motion. Note the similarity between the trajectories of the cusp and the BHs.}
    \label{fig:buoyancy}
\end{figure*}

To gain insight into the origin of dynamical buoyancy, we use the {\tt Buoyancy} simulations, for which we (instantaneously) introduce a BH of mass $\mbh=10^{-3}$ at the center ($r=0$) of different $\abg$ systems (see Table~\ref{table:sims}). We have experimented with alternative methods of introducing the BH into the system, including growing it from a seed mass and dropping it in after evolving the system in isolation for a period of time, and found indistinguishable results. The left panel of Figure~\ref{fig:buoyancy} shows the resulting motion of the BHs in four systems with $\beta=4$ and $\gamma=0.1$ systems, but with different values of $\alpha$, as indicated. These are the same four systems as in  Figure~\ref{fig:profiles}, which shows that while the density profiles are similar, their DFs reveal large differences. 

In the $\alpha=1$ and $\alpha=2$ systems (which do not have inflections in their DFs), the BH only executes Brownian motion without exhibiting any large-scale motion. This Brownian motion is due to discreteness of the N-body system and hence is a numerical artifact. Increasing the simulation resolution decreases the Brownian motion amplitude \citep{Merritt2007, Gualandris2008}. 
    
In the $\alpha=3$ system, the BH is slowly pushed outward from the center. Over time it starts to move along a fairly eccentric orbit that takes it out to a distance of $\sim 0.1$ from the system's center of mass. The $\alpha=4$ system, which has a deep trough in its initial DF, shows an extreme case of buoyancy. The BH quickly moves out from the center all the way to $r_{\rm BH} \approx 0.5$. We note that as it moves outward, the BH has an azimuthal frequency of $\Omega_{\rm BH}\approx 0.58$, nearly equal to the pattern speed of the unstable dipole mode predicted from linear response theory ($\Omega_\rmp=0.59$). Subsequently, it rapidly sinks back to $r_{\rm BH}\sim 0.15$ where it stalls along a fairly elliptical orbit. This late-time stalling radius is not the same as $r_{\rm stall}$ from Figure~\ref{fig:inspiral}. 

Note that in the $\alpha=3$ and $4$ cases, even though the DF has an inflection, the gradient in the initial DF at the energy of the BH is still negative (i.e. $\gradf <0$). Hence, this suggests that the motion of the BH is not due to the LBK torque (which is expected to be negative), but rather due to the torque induced by the unstable dipole mode.

The {\tt Buoyancy} simulations of the $\abg=(5,4,0.3)$ system further confirm this picture. This system has a central cusp that is easily measurable using the center of mass of the 500 most dense particles. In addition, its DF has an inflection, and when evolved in isolation, is unstable to the growth of a dipole mode. The torque arising from the dipole mode dislodges the central cusp and sets it into motion. The black curve in the right-hand panel of Figure~\ref{fig:buoyancy} shows the trajectory of the cusp. The subsequent curves show the motion of BHs of different mass, as indicated, in simulations of the same host system in which the BH is instantaneously introduced at rest at the center of the system at $t=0$. Note that the cusp and the BHs show remarkably similar trajectories: they initially move outward to a maximum radial distance of $r_{\rm BH} \approx  0.4$, then 'turn around' and sink down to $r_{\rm BH} \approx 0.1$, where they continue to orbit for a long time. By tracking the orbits of the individual particles that initially made up the cusp, we have verified that the BH sits atop the cusp as it moves outward. This strongly suggests that the dipole instability is the underlying mechanism for the outward motion of the BH, and thus for the buoyancy seen in these simulations. The time lag between the $\mbh=10^{-3}$ case  and the cusp and $\mbh=10^{-4}$ cases reflects the fact that the massive BH modifies the system's DF, which impacts the growth of the dipole mode \citep[see also][]{vdBosch.Dattathri.25}.

Interestingly, these findings give a natural explanation for the apparent discrepancy between the results of \citet{Cole2012} and \citet{Meadows2020}. The former were the first to identify dynamical buoyancy in a suite of simulations of cored systems. However, the physicality of buoyancy was subsequently questioned by \citet{Meadows2020}, who failed to detect any sign of buoyancy in their simulations. While \citet{Cole2012} modeled their host system using a double-power law density profile with $\abg=(3.7,4.65,0.07)$, \citet{Meadows2020} used a non-singular isothermal density profile. The former has an inflection in its DF, while the latter doesn't. Hence, if buoyancy arises from the dipole instability, which in turn requires an inflection in the DF, this naturally explains the discrepancy between these studies.

\section{Discussion}
\label{sec:discussion}

We have shown that the local theory of dynamical friction (equation~[\ref{eq:c43}]) breaks down in systems with a shallow inner density slope, giving rise to unexpected dynamics: core stalling and dynamical buoyancy. Furthermore, it is clear from figures~\ref{fig:inspiral} and \ref{fig:buoyancy} that not all cores give rise to this dynamics. In particular, the motion of the BH depends crucially on the overall shape of the host's DF, which can drastically differ amongst systems with similar density profiles. 

In this section, we discuss the physical mechanisms underlying friction, stalling, and buoyancy. We also compare our findings with existing models in literature, and discuss some outstanding questions raised by our framework.  

\subsection{Role of DF gradient}
\label{ssec:gradient}

We have shown that the gradient in the DF is the key determinant governing core dynamics. Here we discuss its role in both the linear and non-linear regimes.

\subsubsection{Linear regime}
\label{sec:linear}

In the linear regime, the motion of an infalling BH is mostly determined by the LBK torque, which predicts that dynamical friction arises from discrete resonances in the system between the field particles and the BH. At each resonance, the gradient of the DF in action space dictates the direction in which energy and angular momentum is transferred. When the orbital frequencies are sufficiently non-degenerate, the resonances form a continuum, i.e. a large range of ($l_1,l_2,l_3$) contribute. In this regime, the LBK torque reduces to the Chandrasekhar torque \citep{Tremaine1984,Weinberg1986}, which is why equation~(\ref{eq:c43}) accurately reproduces the BH's trajectory at initial times (Figure~\ref{fig:inspiral}).

On the other hand, if the inner density slope $\gamma$ is small, the orbital frequencies become degenerate within the core, and hence the continuum assumption is no longer valid \citep{Weinberg1986, Read2006}. As shown by \citet{Kaur2022}, the total torque on the BH is then dominated by the contribution from the corotation resonance. The magnitude of the corotation resonance torque depends on $\gradf$. Physically, $\gradf$ dictates the relative number of particles that gain {\it vs.} lose energy to the BH. When this gradient vanishes, there are an equal number of losers versus gainers, and the BH stalls as there is zero net torque acting on it.

As discussed in Section~\ref{sec:background}, in deriving the expression for the LBK torque the self-consistent response of the system, which includes its own self-gravity, has been ignored. However, if the system itself is unstable (e.g. if its DF has an inflection), then the resulting instability can have an important impact on the perturber. This is the case in the {\tt Buoyancy} simulations shown in Section~\ref{sec:buoyancy}; the inflection in the DF renders the system unstable to a dipole mode, which exerts a net torque on the BH and dislodges it from the center. The resulting outward motion of the BH is a manifestation of dynamical buoyancy. 

\subsubsection{Non-linear regime}
\label{sec:nonlinear}

As the BH approaches the stalling radius, it enters the ``slow regime'', in which non-linear effects such as orbit trapping become relevant. However, as is evident from Figure~\ref{fig:dist_func_plots}, the $\rmd f/\rmd E = 0$ criterion for stalling still seems to be applicable in this slow regime, as long as the perturber is not too massive. In this case, the size of the libration zone remains small compared to the size over which the DF changes significantly, and the libration zone itself evolves negligibly over time. Phase mixing among the librating orbits will drive their contribution to the net torque to zero after a few libration periods, which explains how stalling can, under the right circumstances, last for indefinitely long periods.

On the other hand, if the perturber is more massive, the libration zone is larger and the BH affects particles over a larger area of phase-space. Since the condition $\rmd f/\rmd E = 0$ is only valid over a small patch of phase space, the BH typically continues to experience a net torque, although weak. As a consequence, the libration zone continues to evolve, which implies that particles cross the separatrix, thereby commencing or terminating their libration. Because of these effects, the stalling of a sufficiently massive perturber is never perfect, and it continues to sink inward, albeit at a rate that is much reduced due to the shoulder in the DF. 

\subsection{Evolution in the Distribution Function}
\label{ssec:nonlinear}

When the BH sinks towards the center, it transfers its orbital energy and angular momentum to the particles that make up the host system, thereby modifying the DF of the host system. 

At early times, when the BH is rapidly sinking through the system, it is in the ``fast regime'' and its motion is governed by the LBK torque. Although the BH now sweeps through the resonances fast enough to prevent build-up of non-linearities in the response, the process is inherently time-asymmetric due to the phase-space inhomogeneity of the host system. As a consequence, the net effect on each particle's actions is small but non-zero. Therefore, the corresponding changes to the DF during this phase are also small. This is apparent from Figure~\ref{fig:dist_func_plots}, which reveals very little evolution in the system's DF prior to stalling. 

However, once the BH approaches a plateau in the DF and slows down, non-linear effects become increasingly relevant. Chief among these is orbit trapping, whereby a particle crosses the separatrix associated with the perturber and becomes trapped into libration \citep[e.g.,][]{Henrard1982, Tremaine1984}. Since the trapped particles have different libration frequencies, they phase-mix over time if the libration zone does not evolve significantly over time (which is true during stalling of low-mass perturbers). Importantly, the orbit can become untrapped when it crosses the separatrix again, either because of some chaotic behavior or because the separatrix itself evolves. Separatrix crossing, and the associated phase-mixing, is a non-reversible effect, resulting in large changes in the system's DF (\citealt{Chiba2022}; see also \citealt{Kodama2026}). For example, in the {\tt RapidCore} system, non-linear effects become relevant once the BH reaches the stalling radius, where $\gradf \approx 0$. Now, orbit trapping works towards erasing the inflection in the DF, beyond what the dipole instability is able to do (cf. Figures~\ref{fig:fE_nobh} and~\ref{fig:dist_func_plots}). Non-linear effects can also become important when the BH has sunk in to a radius where its mass becomes comparable to the enclosed mass. An example of this is the sinking of the BH in the {\tt GradualCore} system shown in the right-hand columns of Figure~\ref{fig:dist_func_plots}. Here, the initial DF does not have any inflection or plateau that would give rise to stalling. However, once the BH reaches $r_{\rm BH} \simeq 0.21$, the enclosed mass becomes comparable to the BH mass and non-linear effects, including orbit trapping works towards {\it creating} a plateau in the DF, which subsequently causes the perturber to stall. Over time, this newly created inflection is eroded over time and the BH begins to slowly sink in again once the DF has evolved sufficiently such that $\gradf<0$. 

\subsection{Comparison with previous studies}
\label{sec:comparison}

Core stalling was first predicted by \citet{Weinberg1986}, who used the \citet{Tremaine1984} model of dynamical friction to argue that the LBK torque will vanish in harmonic cores with degenerate frequencies, giving rise to stalling. This was subsequently studied by \citet{Read2006} using both semi-analytic and full N-body simulations. They suggest that stalling is a property exclusive to perfectly constant density (harmonic) cores, and hence should only be observable in density profiles with $\gamma=0$. However, it is clear from Figure~\ref{fig:inspiral} that this is not the case.


In order to explain stalling in systems that do not (initially) have a central core, several works have proposed that as a massive perturber sinks in, it dynamically heats the central region and creates a core, resulting in stalling. A commonly used criterion \citep[][]{Goerdt2010, Chowdhury2019, Petts2015, DiCintio2025} is that stalling occurs at the radius where the enclosed mass is comparable to the BH mass, i.e., $M_{\rm enc}(r_{\rm stall})\sim M_{\rm BH}$. Similarly, \citet{Petts2016} argue that stalling occurs when the perturber tidally disrupts the central cusp, and therefore the stalling radius is equal to the tidal radius. Although these models have seen some success, they cannot accurately reproduce all the stalling radii in Figure~\ref{fig:inspiral}. For example, in the {\tt RapidCore} system with a BH of mass $\mbh=10^{-3}$, the enclosed mass radius and the tidal radius are equal to $0.17$ and $0.31$, respectively; both are significantly smaller than the observed stalling radius of $0.65$. In addition, both the enclosed mass model and the tidal radius model predict that the stalling radius increases with perturber mass. This may be the case if the system's DF initially has no inflection or plateau, since a more massive perturber will induce a stronger response, leading to larger changes in the DF. However, in systems that already have a plateau in the DF, such as the {\tt RapidCore} system, perturbers stall at the same radius independent of their mass (Figure~\ref{fig:pert_mass}).

\citet{Banik2022} formulate a non-perturbative treatment of dynamical friction by treating dynamical friction as a restricted three-body problem. By integrating orbits that are close to resonances with the perturber, they show that outside the core, the near-resonant orbits exert a net negative torque on the perturber, resulting in friction. However, inside the core, following a bifurcation in the Lagrange points, some of the near-resonant orbits exert a net positive torque, resulting in buoyancy. Hence, they argue that stalling should occur at or near this bifurcation radius, $r_{\rm bif}$ \citep{Banik2023}. Importantly, bifurcation strictly only occurs in perfectly harmonic cores ($\gamma=0$). For the two $\gamma=0$ profiles considered here, for $M_{\rm BH}=10^{-3}$, the $\alpha=1$ system has $r_{\rm bif}=0.194$, and the $\alpha=2$ system has $r_{\rm bif}=0.368$, which are slightly larger but in reasonable agreement with the $N$-body results in Figure~\ref{fig:inspiral}. The exact role of bifurcation in stalling and buoyancy is left for future work.

\citet{DiCintio2025} suggest that core stalling is due to force fluctuations induced by discreteness effects. Using a series of full N-body and tracer particle simulations, they find a $N^{-1/2}$ scaling of the stalling radius with resolution, which they interpret as evidence for granularity-induced stalling (similar to Brownian motion). However, it is crucial to note that the ``stalling radius'' defined by \citet{DiCintio2025} refers to the radius where a perturber orbits indefinitely without further infall. This differs from our definition, which is the radius at which the trajectory of a perturber suddenly deviates from the Chandrasekhar prescription. Unlike discreteness-induced stalling, the stalling studied in this paper is a physical collective effect due to the vanishing of gradients in the DF. In particular, as explicitly shown in Appendix~\ref{sec:appendix}, it is independent of numerical resolution.

In this regard, we emphasize that we have shown that systems with very similar density profiles (differing only in their $\alpha$ values) host markedly different BH trajectories. It is therefore clear that any theory for stalling based solely on the system's density profile would be incomplete. A full explanation necessarily involves kinetic theory. \citet{Kaur2018} directly evaluate the LBK torque in a cored isochrone system, for which all actions are analytic. They find that within the core, the number of resonances as well as the torque due to each resonance rapidly diminish within a ``filtering radius''. Similar to the bifurcation radius of \citet{Banik2022}, the filtering radius is only defined for constant density cores ($\gamma=0$). For $M_{\rm BH}=10^{-3}$, the filtering radii for our two cored system are $r_{\rm fil}=0.144$ for $\alpha=1$ and  $r_{\rm fil}=0.292$ for $\alpha=2$ system. These values are in slightly better agreement with our simulation results than the bifurcation radii discussed above.

Later, \citet{Banik2021} modified the LBK formula to relax the adiabatic and secular approximations, and found an additional contribution from a ``memory effect'', which can result in an outward torque (buoyancy) within the filtering radius. It is unclear whether this is the same type of dynamical buoyancy as that discussed in Section~\ref{sec:buoyancy}. In particular, in all cases studied here buoyancy seems to be directly associated with the onset of a dipole instability, with no reference to any memory effect. Whether the buoyancy envisioned by \citet{Banik2021} is perhaps a separate effect from that studied here is an open question that we leave for future work.

Finally, a recent study by \citet{Boily2025} examines the migration of BHs from the centers of host galaxies with a net sense of rotation, that is, systems with a DF of the form $f(E,L_z)$. They find that if the host is dynamically warm, the BH can gain angular momentum, leading to outward motion akin to our dynamical buoyancy results (Figure~\ref{fig:buoyancy}). \citet{Boily2025} refer to this as ``dynamical traction''. Furthermore, in the case of dynamically cold systems, the system quickly fragments into stellar substructures and clumps, which can significantly delay the BH's migration to the galactic center, or prevent it from remaining at the center. These mechanisms seem to be different from the dynamical buoyancy discussed here. In particular, our host systems are assumed to be isotropic spheres with zero net angular momentum, in which buoyancy arises from collective effects arising from the unstable dipole mode. It remains to be seen whether traction and buoyancy share physical commonalities or whether these are two independent mechanisms that can both dislodge a massive BH from the center of its host and set it in motion. 

\subsection{Open questions} 
\label{sec:questions}

Despite its success in accurately describing the N-body results, the seemingly simple notion that $\gradf$ determines when and where stalling occurs is subject to a number of caveats and open questions. 

Firstly, as evident from the LBK torque formula (equation~[\ref{eq:LBK_isotropic}]), the total torque on the BH depends on the gradient of the DF at {\it all} resonances. For the case of a circular orbit near a central core, \citet{Kaur2018} and \citet{Kaur2022} have shown that the dominant resonance is co-rotation, in support of the $\gradf$ picture advocated here. Indeed, we have analyzed the phase-space changes in the {\tt RapidCore} system as the BH sinks in, and found that the particles surrounding the corotation resonance undergo $\gtrsim 12$ times greater change in energy than those surrounding other resonances. The contribution from the other resonances in an isochrone sphere was analyzed by \citet{Kaur2022}, who found that they can contribute a fraction of the Chandrasekhar torque, preventing complete stalling. It is unclear whether the same mechanism operates in generalized $\abg$ models. Given that the {\tt RapidCore} system shows near-perfect stalling for a prolonged period of time, we suspect that, at least in this case, non-corotation resonances do not play a significant role.  

Relatedly, as shown in Appendix~\ref{sec:appendix}, the notion that $\gradf$ dictates stalling also appears to hold in the case of eccentric orbits. This suggests that co-rotation remains the dominant resonance even for highly eccentric orbits, which is a surprising result that merits further study.

Finally, we note that the whole premise of linking the torque to the gradient of the DF in energy is based on the assumption that the system is isotropic and remains so as the BH spirals inward. Generally, realistic galaxies and their dark matter halos are triaxial with anisotropic dynamics. Studying whether and how stalling and buoyancy operate in such systems is left for future work.

\section{Astrophysical implications}
\label{sec:implications}

Since cores seem to be common, both in galaxies and in their dark matter halos, the results presented here have a wide range of important implications.

\subsection{Timing problem for globular clusters}
\label{sec:timingproblem}

The presence of multiple globular clusters in several dwarf galaxies has been used as an argument for dark matter cores. This `timing problem' is especially prominent in the Fornax dwarf spheroidal  \citep{Goerdt.etal.06, Cole2012, Boldrini2019, SanchezSalcedo2022}, and recent works have also extended this to other dwarf galaxies \citep{Chowdhury2019, Bar2022, Modak2023, BenYeda2025}. However, it is clear from this paper that the mere presence of a core, i.e. a central region with $\lim_{r\to 0}\rmd\log\rho/\rmd\log r = 0$, does not necessarily imply stalling. For example, in most of the $\alpha=1$ systems in Figure~\ref{fig:inspiral}, the BH is able to quickly sink down to $\sim 0.1 r_s$ (corresponding to $\sim 100$ pc for our dwarf scaling, see Section~\ref{sec:IC}), at which point other processes like dynamical friction from stars and/or gas can take over. More generally, as seen in Figure~\ref{fig:inspiral}, cores with different values of $\alpha$ lead to profoundly different dynamics, due to large differences in their DFs. Although our study here is restricted to spherical isotropic $\abg$ models, our key takeaway, that it is the overall shape of the DF that dictates dynamical friction, is applicable to gravitational systems in general. Hence, this motivates a thorough investigation of the phase-space distribution of cores formed through various mechanisms, such as stochastic outflows, dynamical friction, and alternative dark matter models (see references in Section~\ref{sec:intro}). For example, \citet{Straight2025} show that the cores that form naturally as a consequence of conduction in SIDM halos transition more quickly from the outer to inner density slopes (that is, have larger $\alpha$-values) than those  formed via baryonic processes inside CDM halos. 

\subsection{Lopsidedness and off-centered nuclei in galaxies}
\label{sec:lopsidedness}

As a result of stalling and buoyancy, nuclear star clusters and AGN can be offset from the centers of their host galaxies. Using high-resolution zoom-in cosmological simulations, \citet{Bellovary2021} show that such off-centered BHs with offsets of several hundreds of parsecs are quite common in galaxies with cored density profiles, suggesting that the offsets are not numerical artifacts but rather physical effects due to core dynamics. Observationally, recent surveys find that a large fraction of AGN in dwarf galaxies are off-centered \citep{Reines2020, Mezcua2024}. Since dwarf galaxies are commonly inferred to reside in cored dark matter halos \citep[][]{deBlok2010}, these offsets may well be a consequence of stalling or buoyancy, although we note that there are several other possible explanations. In addition, the dipole instability, and the resulting dynamical buoyancy, can result in lopsidedness in galaxies, which is discussed in detail in \citet{Dattathri2025}.

\subsection{Implications for Black Hole Merger Rates}
\label{sec:BHmergers}

The inspiral and coalescence of massive black holes (BHs) are central to models of galaxy evolution and to predictions of the gravitational-wave sky. Following a galaxy merger, the two central BHs are expected to lose orbital energy via dynamical friction until they form a bound binary, which subsequently hardens through stellar scattering, gas torques, and finally gravitational-wave (GW) emission \citep{Begelman.etal.80, Merritt2013}. However, our results suggest that this canonical picture can be strongly impacted, or even halted, by the phenomena of core stalling and dynamical buoyancy. 

In the case of collisionless systems studied here, when a remnant of merger develops a central constant-density core (pre-existing or produced dynamically during the merger), the inspiral of each BH can stall once it encounters a plateau in the distribution function (DF) of the host system. At this point, the LBK torque vanishes and the frictional drag ceases. As a result, the two BHs may remain separated by tens to hundreds of parsecs, never reaching the regime where stellar hardening or GW emission dominates. Such stalled systems could persist over cosmological timescales, effectively suppressing the formation of close SMBH binaries \citep{Milosavljevic2001, Tamfal2018, DeCun2023}.

The sensitivity of inspiral efficiency to the detailed structure of the underlying DF introduces a new and potentially dominant source of uncertainty in the predictions of SMBH merger rates that are already ridden with uncertainties and an active area of investigation \citep{Grobner+2020, Bortolas2022}. Most semi-analytic and cosmological models assume that following a galaxy merger, their BHs rapidly sink to the center and pair through efficient dynamical friction \citep{Volonteri2009, Barausse2020, Volonteri2020, IzquierdoVillalba2023, Dekel2025}. However, our results demonstrate that this efficiency depends not only on the inner density slope $\gamma$, but crucially on $\gradf$, the DF gradient at the orbital energy of the BH. Systems with inflections or plateaus in their DFs, typical of double power-law profiles with rapid outer-to-inner transitions, are especially prone to stalling or buoyancy. Consequently, merger rates derived from halo merger trees or semi-empirical prescriptions that ignore the fine-structure of the DF of the merger remnant may be systematically overestimated. We emphasize that the dynamics studied here would result in separations of the order hundreds of parsecs, which is still in the regime where dynamical friction is expected to be the dominant mechanism for bringing the BHs together. In particular, it is well outside the regime where stellar hardening (pc scales, \citealt{Quinlan1996}), circumbinary disks (sub-pc scales, \citealt{ArmitagePN2002}), or gravitational wave emission (mpc scales, \citealt{Peters1964}) are expected to be relevant. 

The impact of this dynamical bottleneck propagates directly to gravitational-wave observatories. The planned \textit{Laser Interferometer Space Antenna} (\textit{LISA}; \citealt{AmaroSeoane2017}) aims to detect mergers of BHs with masses $10^{4}$--$10^{7}\,M_{\odot}$ at redshifts $z \lesssim 10$. In this mass and redshift range, most events are expected to arise from mergers within dwarf galaxies or low-mass halos \citep{Barausse2020, Volonteri2020}, precisely the environments where cored profiles and DF plateaus are prevalent. If core stalling is common, a substantial fraction of potential \textit{LISA} sources may never reach the GW regime, lowering predicted event rates relative to current estimates. Indeed, \citet{Tamfal2018} performed high-resolution N-body simulations of dwarf galaxy mergers to study the formation of BH binaries in systems with different DM halo density profiles. They find that in $\abg=(1,4,\gamma)$ DM halos with $\gamma \lesssim 0.6$, the BHs stall at 50-100 pc scales for more than a Hubble time. In light of the results presented here, this stalling can be even more pronounced in systems with larger $\alpha$ or lower $\gamma$ values.

Similar dynamics may also be prominent in massive elliptical galaxies, which are thought to be a hot spot for gravitational wave events \citep{Zhao2025}. A large fraction of massive ellipticals have cores in their stellar density profiles that are characterized by a ``break radius'', where the density sharply transitions to a shallow core, akin to a high $\alpha$ value \citep{Ferrarese.etal.94, Graham2003, Merritt2013, Rawlings2025}. These cores are believed to have formed due to binary BH scouring and/or recoil kicks \citep[][]{Faber1997, Milosavljevic2001, Nasim2021}. Based on the results presented here, an infalling BH in these systems may show  pronounced stalling. In fact, off-center nuclei and 'blobs' have been detected within the cores of such massive ellipticals, which have been suggested to be stalled satellites and/or BHs \citep{Lauer2002, Laine2003, Postman2012}. These galaxies are mostly gas-poor, so gas-driven torques are unlikely to play a significant role. These effects may influence the nanohertz GW background probed by pulsar timing arrays, where the abundance of long-lived, non-coalescing SMBH pairs can imprint a suppression or turnover in the low-frequency strain spectrum.

The dynamical processes discussed in this paper add additional uncertainty to the already large dispersion in current predictions for SMBH merger rates. However, by linking the onset of stalling and buoyancy to the presence of DF plateaus and inflections, it provides a predictive criterion for BH pairing efficiency that can be incorporated into cosmological simulations and semi-analytic models. Future work combining this criterion with high-resolution hydrodynamic simulations will allow us to quantify how baryonic feedback, merger-driven core formation, and dark matter microphysics shape the demographics of stalled and coalescing SMBHs. Ultimately, the gravitational-wave sky may encode not only the cosmic history of black hole growth but also the fine-grained phase-space structure of galaxy cores. 

\section{Conclusions}
\label{sec:conclusions}

Dynamical friction, long treated as a robust and universal drag force, proves far more nuanced than the classic \citet{C43} formalism, which assumes an infinite, homogeneous, and isotropic background medium. These discrepancies are especially prominent in systems with central cores, in which previous studies have shown that dynamical friction ceases to be efficient (core stalling) or even reverse in direction (dynamical buoyancy). However, it has remained unclear what the detailed criteria are under which these phenomena operate.

Our study demonstrates that the efficacy and direction of dynamical friction depend not on the central density gradient alone but on the underlying phase-space structure encoded in the distribution function (DF). Using idealized, high-resolution $N$-body simulations and analytic arguments from kinetic theory, we have shown that core stalling and dynamical buoyancy arise from distinct features in the DF:
\begin{itemize}[leftmargin=0.27truecm, labelwidth=0.2truecm]

  \item \textbf{Stalling} occurs when the inspiraling perturber reaches a plateau in the DF, that is,
  \begin{equation}
    \gradf=(\rmd f/\rmd E)_{E_{\rm BH}}=0 \,,
  \end{equation}
  which causes the net torque to vanish.
    
  \item \textbf{Buoyancy} emerges in systems whose DFs possess an inflection ($\rmd f/\rmd E>0$), rendering them unstable to a growing dipole mode that drives the central mass outward.

\end{itemize}

Crucially, these DF features are not uniquely tied to a specific density slope: systems with similar surface densities or core sizes can behave dramatically differently depending on how sharply their outer and inner density slopes transition. In the case of double power-law $\abg$ models, systems with low $\gamma$ (shallow inner slope) and/or high $\alpha$ (sharp transition) naturally develop plateaus and inflections, explaining why apparently ``similar'' cores yield dissimilar dynamical outcomes. Hence, not all cores are equal.

The DF framework also clarifies why more massive perturbers can transiently overcome the stalling barrier: their inspiral dynamically reshapes the DF, erasing existing plateaus and re-establishing negative gradients that re-enable friction. Conversely, lighter objects remain trapped at the plateau, producing long-lived off-center configurations. In addition, even if the DF initially does not have a plateau or inflection, the BH inspiral itself can create an inflection, resulting in its stalling. This self-regulated interplay between the perturber and its host distribution reveals dynamical friction as a fundamentally nonlinear, feedback-driven process, not a static background drag.

Our results bridge the gap between kinetic theory and cosmological structure formation, offering a unified physical picture of friction, stalling, and buoyancy in cored systems. Future work will extend this framework to more realistic, anisotropic, and time-evolving galaxies, enabling quantitative predictions for the demographics of stalled and wandering black holes across cosmic time. The astrophysical implications are far-reaching. In dwarf galaxies and massive ellipticals, core stalling provides a natural explanation for displaced nuclear star clusters and off-centered AGN. It also accounts for globular clusters in dwarf galaxies which would have otherwise sunk to the center. For SMBH binaries, stalling and buoyancy identify a key bottleneck in the inspiral process prior to binary formation and gravitational-wave emission, influencing predicted merger rates for LISA. This underscores the necessity for an in-depth study of core formation via different mechanisms, focusing in particular on their phase-space distributions, and the resulting dynamics in these systems. One of these mechanisms, self-interacting dark matter, is studied in a follow-up paper \citep{vdBosch.Dattathri.25}.


\section*{Acknowledgments}

We thank the anonymous referees for their careful reading of the paper and useful feedback. This research was supported by the National Science Foundation (NSF) National Science Foundation (NSF) - US-Israel Binational Science Foundation (BSF) through grant AST-2407063 and in part by grants NSF PHY-2309135 to the Kavli Institute for Theoretical Physics (KITP) and AST-2307280 to FvdB. PN gratefully acknowledges funding from the Department of Energy grant DE-SC0017660 and grant 63406 from the John Templeton Foundation. She also acknowledges the Gordon and Betty Moore Foundation (Grant \#8273.01) and the John Templeton Foundation (Grant \#62286) for their support of the Black Hole Initiative. UB acknolwedges funding from the Bezos Membership grant at the IAS. ZZL acknowledges the Marie Skłodowska-Curie Actions Fellowship under the Horizon Europe programme (101109759, ``CuspCore'').


\bibliographystyle{mnras}
\bibliography{references}{}

\appendix
\numberwithin{figure}{section}
\numberwithin{table}{section}
\numberwithin{equation}{section}

\section{Verification and convergence tests}
\label{sec:appendix}

Here, we present several convergence and verification tests to assert the robustness of our results. These constiute the {\tt Convergence} suite of simulations in Table~\ref{table:sims}.

\subsection{Initial radius}
\label{app:init}

Throughout, in all simulations of core stalling, we initialize the BH on a circular orbit at an initial radius $r_{\rm init}=1$, which is equal to the scale radius of the host. As eluded to in the text and discussed in \citet{Banik2021}, the instantaneous introduction of a perturber can introduce transients that may impact the subsequent evolution. In order to ensure that such transients do not have a significant impact on any of our main results, we run two simulations of the {\tt RapidCore} system with $M_{\rm BH}=10^{-3}$, but starting the BH from an initial radii of $r_{\rm init}=2.5$ and $r_{\rm init}=5$. These results are compared against the {\tt Fiducial} simulation, which has $r_{\rm init}=1$ and $r_{\rm stall}\approx 0.65$, indicated by the black dashed line. 

\begin{figure*}
    \centering
    \includegraphics[width=\textwidth]{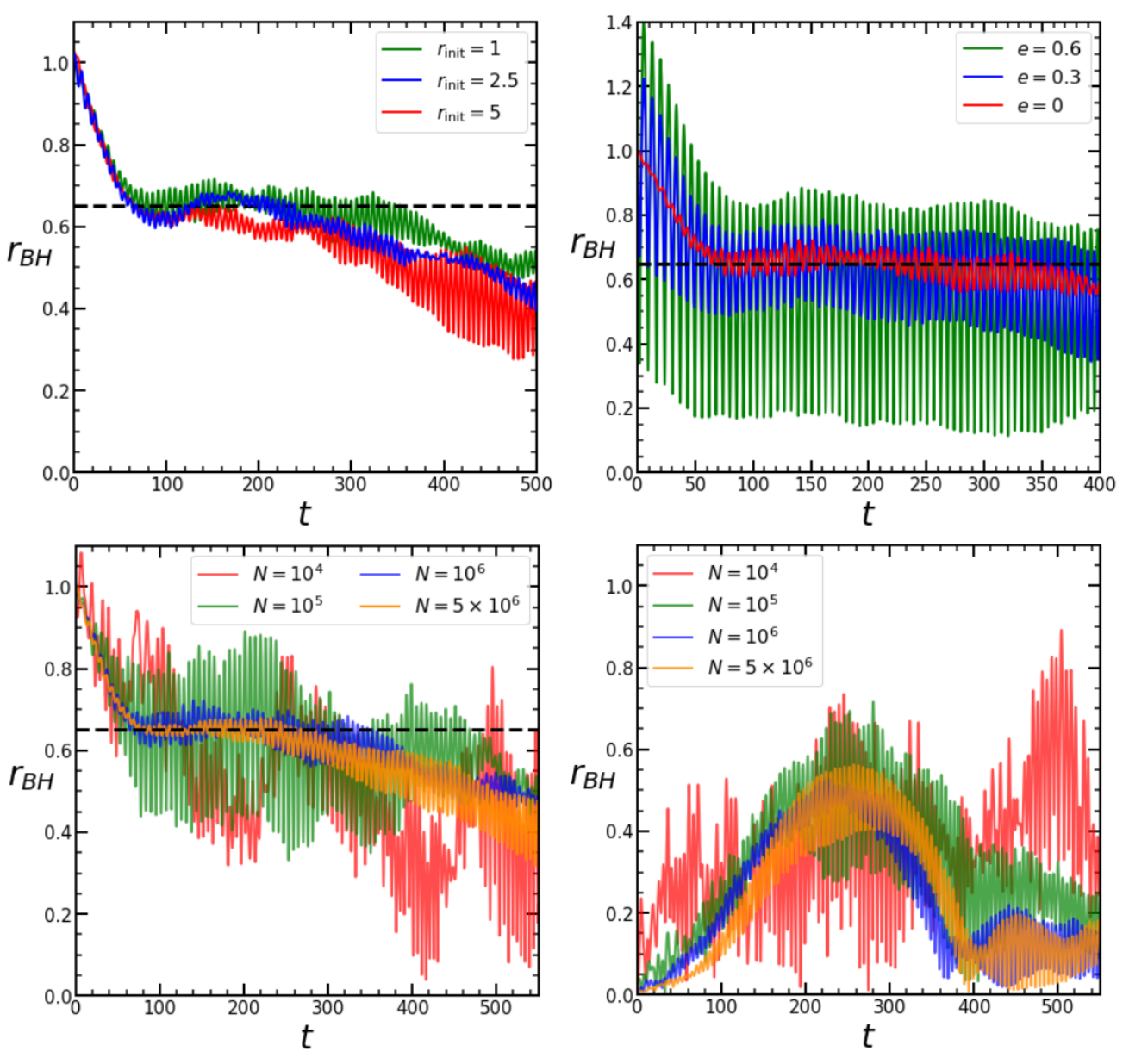}
    \caption{Convergence tests to verify the robustness of our results. In all cases, we plot the trajectory of a BH of mass $\mbh=10^{-3}$ in the {\tt RapidCore} system, i.e. $\abg=(4,4,0.1)$. The horizontal black dashed lines denote the stalling radius $r_{\rm stall}\approx 0.65$ in the {\tt Fiducial} simulation. Top left panel: effect of varying the BH's initial radius $r_{\rm init}$. The trajectories are shifted along the time axis such that $r_{\rm BH}(T=0)=1$ in all cases. Top right panel: trajectories with varying initial eccentricity. From these two panels, we see that the stalling radius is independent of the BH's initial radius and eccentricity. Bottom panels: effect of simulation resolution on stalling (left) and buoyancy (right). A well-resolved trajectory is seen only when $N \geq 10^6$ particles.}
    \label{fig:convergence}
\end{figure*}

The top left panel of Figure~\ref{fig:convergence} shows the trajectory of the BHs in all three cases. In order to better visualize the (lack of) differences between the trajectories, we shift the time axis so that $t=0$ coincides with $r_{\rm BH} =1$. Except for small inconsequential differences at late times, the three trajectories agree well with each other. In particular, they all stall at $t \sim 100$ when they have reached a halo-centric distance of  $\approx 0.65$, followed by a very slow inspiral at late times. We therefore conclude that our results are independent of $r_{\rm init}$, and that transients that arise due to the instantaneous introduction of the BH are negligible \citep[see also][for a similar test, reaching the same conclusion]{vdBosch.Dattathri.25}. We have also experimented with introducing the BH after allowing the system to evolve in isolation for a period of time, in order to ensure that initial transient effects are not important, and again we found no significant differences in the BH trajectory. Therefore, using $r_{\rm init}=1$ is appropriate and saves us significant computational time. 

\subsection{Eccentric orbits}
\label{app:eccentric}

Throughout the paper, we have limited our analysis of core stalling to circular orbits. Here, for comparison, we consider two cases in which the initial trajectory of the BH is eccentric. The top right panel of Figure~\ref{fig:convergence} shows the trajectory of a BH of mass $10^{-3}$ in the {\tt RapidCore} system, with the same parameters as the {\tt Fiducial} simulation, except for the initial orbital eccentricity of the BH. In each simulation, we start the BH from $r_{\rm init}=1$, and with the speed corresponding to the circular velocity at that radius, but with the velocity vector pointing outward by varying degrees, giving rise to different orbital eccentricities. In all cases, the BH stalls when its average orbital radius approaches $\approx 0.65$, similar to the stalling radius in the case of the circular ($e=0$) orbit, which is indicated by the black dashed line. Note that dynamical friction does not cause any net change in orbital eccentricity \citep{vdb1999}. Therefore, BHs with higher initial eccentricity show larger radial oscillations about the stalling radius. 

We have also analyzed the evolution in the DF of the simulations with varying orbital eccentricity (not shown). In all cases, we find that the BH stalls when it reaches the plateau in the DF where $\gradf\approx0$. This suggests that the dominant resonance within the core is corotation, regardless of whether or not the orbit is circular. Hence, our picture of linking stalling to a vanishing $\gradf$ seems to hold even for eccentric orbits.

\subsection{Resolution tests}
\label{app:tests}

The finite resolution of our $N-$body simulations can impact our results in several ways. Firstly, two-body relaxation will cause the diffusion of particles in phase-space, which can impact the torque responsible for dynamical friction \citep[][]{Hamilton.etal.23} and may lead to artificial evolution of the system's DF. In particular, as explicitly demonstrated in \citet[][]{vdBosch.Dattathri.25}, collisions can erase inflections or plateaus in the DF. Since the emergence of stalling and buoyancy critically depends on the presence of such plateaus and inflections, respectively, these phenomena can only be resolved properly if the resolution of the simulation is sufficiently high. Secondly, in addition to core stalling and buoyancy (which are physical effects), the BH also exhibits Brownian motion, which is resolution-dependent and follows a $v_{\rm rms} \sim N^{-1/2}$ scaling. Hence, Brownian motion can dominate over core stalling and buoyancy in low-resolution simulations. 

The bottom panels of Figure~\ref{fig:convergence} show the trajectory of a BH of mass $\mbh=10^{-3}$ in the {\tt RapidCore} system, when inspiraling from $r_{\rm init}=1$ (bottom left panel) and when initially at rest at the center (bottom right panel). The different curves correspond to simulations of different $N$, as indicated. Note how the results for $N=10^4$ (red curves) are extremely noisy, indicating that neither stalling nor buoyancy is adequately resolved at this numerical resolution.  With $N=10^5$ particles, the BH clearly exhibits stalling and buoyancy, but the trajectories are still too noisy for a meaningful analysis, as evident from the erratic radial oscillations. Only with $N \geq 10^6$ particles are the BH trajectories well-resolved. However, these results obviously depend on the mass of the BH.



\label{lastpage}
\end{document}